\documentclass[preprint,12pt]{elsarticle}

\usepackage{amssymb}
\usepackage{amsmath}
\usepackage{graphicx} 
\usepackage{subcaption}
\usepackage{hyperref}
\usepackage{amssymb}
\usepackage{graphicx}
\usepackage{float}
\usepackage{amsmath}
\usepackage{color}
\usepackage{algorithm}
\usepackage{algpseudocode}
\usepackage{booktabs}
\usepackage{multirow}

\journal{Ocean Engineering}

\begin{document}

\begin{frontmatter}



\title{Length-Preserving Vessel Trajectory Simplification With Controllable Compression-Precision Bound}

\author[cosco]{Enmei Tu}
\author[dlmu]{Zhiye Xie}
\author[cosco]{Xixia Xu}
\author[cosco]{Shuhang Chen} 
\affiliation[cosco]{organization={COSCO SHIPPING Advanced Technology Institute}
            }

\affiliation[dlmu]{organization={School of Information Science and Technology, Dalian Maritime University}
            }


\begin{abstract}
Vessel trajectory data from the Automatic Identification System (AIS) is important for a wide range of maritime analysis and management applications. As AIS datasets continue to grow in scale, trajectory simplification (or compression) algorithms have become indispensable tools for managing their computational demands. Existing approaches, however, largely overlook the preservation of trajectory length, a critical metric of direct relevance to maritime environmental auditing and voyage management. We address this gap by framing trajectory simplification as a pair of dual optimization problems subject to error constraints: Compression-Driven Simplification (CDS), which minimizes data volume for a given length-error tolerance, and Precision-Driven Simplification (PDS), which minimizes length approximation error for a given compression target. Both problems are solved within a unified dynamic programming framework, and we provide mathematical proofs establishing that our formulation yields globally optimal solutions. This guarantees flexible, direct control over both compression ratio and length-approximation fidelity. { We benchmark CDS and PDS on real-world AIS datasets spanning diverse navigation scenarios and representative water area, comparing against state-of-the-art baselines. Experimental results on public benchmark datasets demonstrate that even under a small length error tolerance ($\epsilon=0.1$ nmi), our methods achieve about 91\% data compression rate,  approximately 14\% higher than the best-performing baseline, which implies that our methods are capable of maximally preserving the trajectory length while reducing data redundancy significant. } 
\end{abstract}
 


\begin{keyword}
Vessel Trajectory Simplification \sep Length Preservation \sep Dynamic Programming \sep AIS Data



\end{keyword}

\end{frontmatter}


\section{Introduction}
The global shipping industry plays a pivotal role in international trade, yet it is also a significant contributor to greenhouse gas (GHG) emissions. In response to growing environmental concerns, the International Maritime Organization (IMO) has implemented strategies to reduce CO$_2$ emissions from shipping, targeting at least a 40\% reduction by 2030 and 70\% by 2050 compared to 2008 levels \cite{IMO2020}. To monitor and verify these reductions, systems such as the EU Monitoring, Reporting and Verification (MRV) and the IMO Data Collection System (DCS) have been established, which rely heavily on accurate data regarding vessel movements and fuel consumption.

Automatic Identification System (AIS) data, providing real-time spatiotemporal information about vessel trajectories, has become indispensable for bottom-up emission modeling and various maritime applications, including traffic management, safety, and energy efficiency analysis. However, the sheer volume of raw AIS data presents significant challenges. Billions of AIS messages are generated daily, leading to substantial storage and computational demands. For instance, modeling fuel consumption for 50,000 ships over a year can require approximately 20,000 CPU hours \cite{Balcer2024}. To mitigate these challenges, trajectory simplification algorithms are frequently employed to reduce data size while preserving essential trajectory characteristics.

\begin{figure}[t]
\centering
\includegraphics[width=0.8\textwidth]{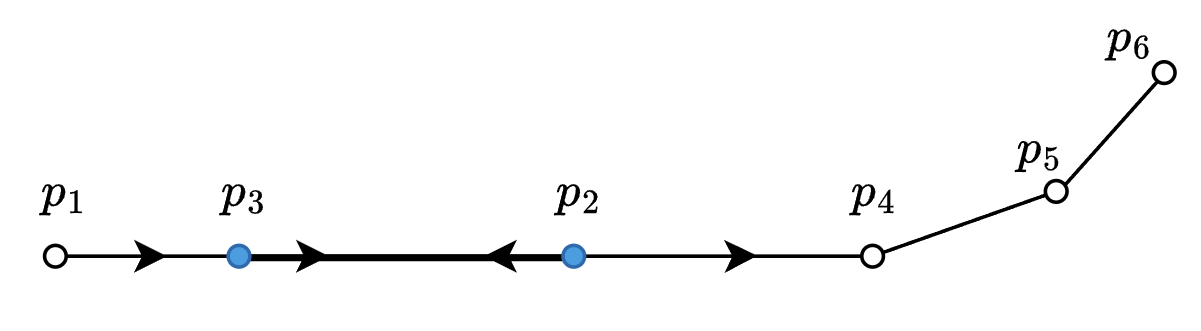}
\caption{Back-and-forth sailing trajectory. The vessel sails from point $p_1$  to  $p_2$ , turning back at $p_2$ to $p_3$, and then reversing again at $p_3$ toward $p_4$.  The blue dots are turning points.}\label{back_forth_fig}
\end{figure}

Existing trajectory simplification approaches, such as the widely used Ramer-Douglas-Peucker (RDP) algorithm and its variants \cite{Douglas1973, zhang2016ais, ye2024adaptive}, primarily focus on preserving the geometric shape of the trajectory. While effective for visual representation and certain spatial analyses, these methods often introduce significant errors in critical metrics like trajectory length, and lack explicit control on error bound.  For example, the back-and-forth sailing trajectory illustrated in figure \ref{back_forth_fig}. For these methods, the journey $p_2\rightarrow p_3 \rightarrow p_2$ tends to be removed easily without changing the geometric shape. For applications such as calculating vessel energy efficiency (e.g., via the Energy Efficiency Operational Indicator, EEOI) \cite{prastowo2025implementation}, estimating fuel consumption \cite{simonsen2018model}, auditing carbon emissions \cite{greene2020well}, and optimizing voyage performance \cite{wang2019three}, the trajectory length is an important input to distance-dependent fuel-consumption and emission estimation. Consequently, simplification methods that do not prioritize length preservation can lead to inaccurate estimations of these vital environmental and operational indicators \cite{Balcer2024}. 

This paper addresses the limitations of current trajectory simplification techniques by proposing a novel dynamic programming (DP) based approach designed to explicitly preserve trajectory length. Specifically, we formulate the problem as a pair of dual optimization problems : 
\begin{itemize}
    \item  \textbf{Compression-Driven Simplification (CDS)}: For a user-specified target number of points $K$, find the simplified trajectory that minimizes the overall length reduction error.
    \item \textbf{Precision-Driven Simplification (PDS)}: For a user-specified maximum allowable length error tolerance $\epsilon$, find the simplified trajectory that minimizes the number of points required to stay within that tolerance.
\end{itemize}
This dual formulation can be solved within a unified dynamic programming framework, allowing for flexible application based on specific data reduction or error tolerance requirements. Through this approach, we aim to provide a simple but effective method for controlling AIS trajectories length that ensures high fidelity in length-dependent applications, thereby enhancing the accuracy of vessel energy efficiency and carbon emission assessments. We validate our approach using real-world AIS datasets, demonstrating its superiority over classic shape-oriented algorithms like RDP in maintaining length accuracy, and its effectiveness in supporting reliable vessel efficiency analysis.

The rest of this paper is organized as follows. Section 2 reviews related work on trajectory simplification. Section 3 formally defines the problem and details our proposed dynamic programming framework for the dual optimization problems. Section 4 presents the experimental results and analysis. Finally, Section 5 concludes the paper and discusses future work.

\section{Related Works}
The proliferation of Automatic Identification System (AIS) data has necessitated the development of efficient trajectory simplification techniques to manage the immense volume of information while preserving crucial characteristics. The existing literature on trajectory simplification can be broadly categorized based on their underlying methodologies and the specific features they aim to preserve. This section provides a comprehensive overview of these approaches, highlighting their strengths and limitations, particularly concerning length preservation. A more comprehensive review can be found in \cite{liu2024ais}.

Geometric-based algorithms are among the earliest and most widely adopted methods for trajectory simplification. The Ramer-Douglas-Peucker (RDP) algorithm \cite{Douglas1973} is a classic example, operating by recursively dividing a trajectory and removing points that fall within a specified perpendicular distance tolerance from the line segment connecting the start and end points. While effective in reducing data volume and preserving the visual appearance of a trajectory, RDP is very sensitive to its distance threshold parameter. To address this problem, adaptive RDP (ARDP) algorithms are developed to determine the threshold automatically based on the trajectory shape \cite{Zhang2024, tang2021method}. However, both RDP and ARDP often fall short in preserving kinematic properties, such as the actual length of the path or the vessel's speed profile. This limitation becomes particularly critical in applications where the precise distance traveled is a key input, such as fuel consumption and emission calculations \cite{Balcer2024}. Other geometric methods like  Visvalingam-Whyatt \cite{Visvalingam1993} also prioritize shape over other characteristics.

To address the shortcomings of purely geometric approaches, kinematic-aware algorithms have emerged. These methods integrate temporal and motion-related information into the simplification process. For instance, the Top-Down Time Ratio (TD-TR) algorithm considers the time dimension during simplification, aiming to better represent the vessel's movement dynamics \cite{Keogh1998}. More recent works, such as the Top-Down Kinematic Compression (TDKC) algorithm proposed by Guo et al., focus on adaptively compressing trajectories while preserving features related to vessel kinematics \cite{Guo2024}. Similarly, Wei et al. introduced an AIS trajectory simplification algorithm that considers ship behaviors, aiming to retain important motion characteristics \cite{Wei2020}. While these methods offer improvements over purely geometric approaches, they may still not explicitly optimize for length preservation, often treating it as a secondary outcome of preserving other kinematic features.

For scenarios requiring immediate processing of continuous AIS data streams, online/streaming algorithms are essential. Methods like the Opening Window (OPW) algorithm \cite{Meratnia2004} and SQUISH-E (Muckell et al., 2011) \cite{Muckell2011} process trajectory points sequentially, making decisions about point retention or removal without requiring the entire trajectory to be available. These algorithms are vital for real-time applications such as collision avoidance systems or dynamic traffic management. However, the incremental nature of these algorithms can sometimes lead to suboptimal global simplification results compared to offline methods, and their primary focus is often on real-time responsiveness rather than precise length preservation.

Optimization-based algorithms, particularly those leveraging dynamic programming (DP), offer a powerful framework for achieving optimal simplification under specific criteria. Long et al. explored direction-preserving trajectory simplification using DP, demonstrating its ability to provide quality guarantees in terms of position-related properties \cite{Long2013, Long2014}. These approaches can be designed to minimize a defined error metric or to achieve a target compression ratio while adhering to certain constraints. Hu et al. also proposed a three-dimensional compression of ship trajectories using a dynamic programming algorithm \cite{Hu2024} by minimizing overall synchronous Euclidean distance between the original trajectory and the simplified trajectory. The strength of dynamic programming lies in its ability to explore a vast solution space to find a globally optimal solution, making it highly suitable for problems where specific properties, such as trajectory length, need to be precisely preserved. Therefore, it provides a strong foundation for our proposed approach, which explicitly formulates length preservation as a dual optimization problem solved efficiently within a unified dynamic programming framework.

{
More recently, several works have proposed further refinements to data-driven and behavior-aware simplification. \cite{li2025simplification} addresses a specific limitation of standard geometric methods by introducing a classification-based strategy tailored to simplifying circular/looping ship trajectories, which conventional threshold-based algorithms tend to distort. \cite{he2026improved} improves AIS trajectory simplification by explicitly considering motion continuity, aiming to produce simplified routes more suitable for typical-route extraction.  \cite{mehri2025novel} propose a context-aware variant of Douglas-Peucker (CADP) that adapts the simplification threshold to local trajectory context rather than applying it uniformly. While these methods improve shape- and context-sensitivity of simplification, none of them explicitly formulates trajectory length preservation as the optimization objective.

\section{Methodology}
In this section,  we first provide a concise overview of AIS data and the dynamic programming technique to establish essential background for unfamiliar readers. Then, we formulate our length-preservation trajectory simplification task as a pair of constrained optimization problems and provide a unified framework to solve them with dynamic programming technique. 
}
\subsection{Preliminaries}

\textbf{AIS Data.} The Automatic Identification System (AIS) is a pivotal tracking technology used in maritime navigation and traffic management. It operates on the VHF band, enabling vessels to automatically and periodically broadcast dynamic information (e.g., position, speed, course, heading) and static information (e.g., Maritime Mobile Service Identity (MMSI), vessel type, dimensions). An AIS message thus can be denoted as $$p=(t, \text{MMSI, LAT, LON, SOG, COG, Heading}, ...)$$ where $t$ is the timestamp, MMSI is the vessel's identifier, (LAT, LON) denotes the geographic coordinates (latitude, longitude), SOG is the speed over ground,  COG is the course over ground and Heading is the direction the ship's hull is pointing.The widespread adoption of AIS has generated vast, high-frequency spatiotemporal data streams, creating unprecedented opportunities for maritime research, including traffic pattern analysis, anomaly detection, route optimization, and environmental monitoring. While AIS data is invaluable for maritime traffic analysis, it possesses several inherent characteristics that necessitate sophisticated processing techniques. 1). AIS data exhibits irregular temporal spacing due to dynamic reporting rates, vessel maneuvers, and transmission conditions. Consecutive points may be separated by seconds to minutes, even to hours and days. 2). During straight-line navigation, consecutive points are nearly collinear, creating substantial data redundancy. A vessel traveling at constant speed and heading generates numerous points that convey minimal additional trajectory information.
3). AIS data are usually very noisy. It contains errors caused by GPS inaccuracy, signal interference, and spoofing. 4). Global AIS networks generate billions of messages daily. A single vessel on a trans-oceanic voyage may produce 10,000+ position reports, necessitating trajectory compression for storage and analysis efficiency.

\textbf{Dynamic Programming.} Dynamic Programming (DP) is a systematic algorithmic paradigm for solving complex optimization problems by breaking them down into simpler, overlapping subproblems. The core principle involves solving each subproblem only once, storing its solution, and reusing these solutions to construct the overall optimal solution. This approach contrasts with naive recursive methods that repeatedly solve identical subproblems, leading to exponential time complexity. More specifically, let $s$  represent a state (configuration) of the problem and $dp[s]$  store the optimal value for that state. Then we can use the following recurrence relation to compute $ dp[s] $ from smaller states:

$$
dp[s] = \begin{cases}
\text{initial\_value} & \text{if } s \text{ is a base state} \\
\min_{a \in A(s)} \left\{ \text{cost}(s, a) \oplus dp[\text{next}(s, a)] \right\} & \text{otherwise}
\end{cases}
$$
where $ A(s) $  is the set of valid actions from state  $s$ , $\text{next}(s, a) $ is the resulting state after action  $a$ , and  $\oplus$  is an operator (typically + for additive costs). The base state is a user predefined state providing the initial value for the simplest and smallest subproblems.

\subsection{Problem Formulations}
Given an AIS data vessel trajectory $\mathcal T$ of length $N\ge 3$, represented as a sequence of points $\mathcal {T} = (p_1, p_2, \ldots, p_N)$, where each point $p_i \in \mathbb{R}^d$ (e.g., $d=2$ for latitude-longitude coordinates). Let $f: \mathbb{R}^d \times \mathbb{R}^d \to \mathbb{R}_{\ge 0}$ be a distance function, typically earth distance computed with Haversine formula. The total length (sailing distance) of $\mathcal T$ is:

\begin{equation}
L(\mathcal T) = \sum_{j=1}^{N-1} f(p_j, p_{j+1})
\end{equation}

A simplified trajectory $\mathcal S=(p_{i_1}, p_{i_2},
\cdots, p_{i_k})$ is a subsequence of $\mathcal T$ of length $k\le N$, represented by an index set $\mathcal I = \{i_1, i_2, \ldots, i_k\} \subseteq \{1,2,\ldots,N\} $ , where $1 = i_1 < i_2 < \cdots < i_k = N$. The number of points in $\mathcal S$ is denoted by $\left|\mathcal I\right|=k$, where $\left|\cdot\right|$ represents set capacity. The total length of the simplified trajectory is:
\begin{equation}
L(\mathcal S) = \sum_{m=1}^{k-1} f(p_{i_m}, p_{i_{m+1}})
\end{equation}

{Define the length difference between the original and the simplified trajectories as:
\begin{equation}
\Delta(\mathcal S) = \left| L(\mathcal T) - L(\mathcal S) \right|
\end{equation}
Since $f$ is a metric (e.g., Euclidean or Haversine distance), it satisfies the triangle inequality: for any three points $p_a, p_b, p_c$, $f(p_a, p_c) \leq f(p_a, p_b) + f(p_b, p_c)$. Applying this repeatedly to each original sub-path between two consecutive retained points $(p_{i_m},p_{i_{m+1}})\in \mathcal{S}$ shows that the direct chord is never longer than the original sub-path:
$$
f(p_{i_m}, p_{i_{m+1}}) \;\leq\; \sum_{j=i_m}^{i_{m+1}-1} f(p_j, p_{j+1}).
$$
Summing this inequality over all segments of $\mathcal{S}$:
\begin{equation}
	\label{delta_s}
L(\mathcal{S}) = \sum_{m=1}^{k-1} f(p_{i_m}, p_{i_{m+1}}) \;\leq\; \sum_{m=1}^{k-1}\sum_{j=i_m}^{i_{m+1}-1} f(p_j, p_{j+1}) = L(\mathcal{T})
\end{equation}
which gives $L(\mathcal S) \leq L(\mathcal T)$. This bound holds for any fixed index set $\mathcal{I}$ and any distance function $f$ satisfying the triangle inequality, independent of trajectory shape or navigation complexity.


}
Based on these definitions, our length preservation trajectory simplification problem can be formulated as an optimization problem as follows:

\begin{enumerate}
	\item[I] \textbf{Precision-Driven Simplification (PDS).}
	 Minimize the number of points in the simplified trajectory, subject to the constraint that the length difference does not exceed a given threshold $\epsilon > 0$.
{
	\begin{equation}
	\begin{aligned}
	\min_{\mathcal I \subseteq \{1,2,\ldots,N\}} \left|\mathcal I\right|\\
	\text{s. t. } \Delta(\mathcal S) \leq \epsilon
	\end{aligned}
	\end{equation} }
	PDS controls trajectory length error within an allowable tolerance while minimizing data points.
	\item[II] \textbf{Compression-Driven Simplification (CDS).}
	 Minimize the length difference, subject to the constraint that the number of points in the simplified trajectory does not exceed a given integer $K$ (where $2 \le K \le N$). 
{
\begin{equation}
	\begin{aligned}
\min_{\mathcal I \subseteq \{1,2,\ldots,N\}} \Delta(\mathcal S)\\
\text{s.t.}
\left|\mathcal I\right|=k \leq K
\end{aligned}
\end{equation}}
CDS controls the number of data points within an allowable amount while minimizing trajectory length error.
\end{enumerate}
Trajectory simplification involves a fundamental trade-off between approximation error and compress ratio \cite{prastowo2025implementation}. Traditional methods, such as the Ramer-Douglas-Peucker (RDP) algorithm and its variants \cite{zhao2018method, lee2022ais}, lack mechanisms to explicitly manage this trade-off. In contrast, the proposed optimization framework provides flexible, direct control over both simplification error and data volume. 
{
For application scenarios, the choice between PDS and CDS should be guided by point budget or length accuracy. CDS is appropriate when downstream storage, transmission, or processing capacity imposes a hard cap on the number of points per trajectory, and it is particularly effective for long, highly redundant trajectories dominated by straight-line sailing, where a modest point budget can still capture almost all of the sailed distance because redundant near-collinear points contribute little length information. PDS, in contrast, is preferable when a specific length-accuracy requirement must be guaranteed, where the acceptable error tolerance $\epsilon$ is dictated by the downstream metric rather than by storage considerations. PDS is especially advantageous for trajectories with frequent maneuvers, sharp turns, or back-and-forth segments, since it automatically allocates more points to geometrically or kinematically complex regions and fewer to simple ones, avoiding the under- or over-compression that a single fixed $K$ would cause across trajectories of heterogeneous complexity.}

{
From equation \ref{delta_s}, we know that $L(\mathcal S) \leq L(\mathcal T)$ holds for any distance function $f$, so $\Delta(\mathcal{S})=L(\mathcal T)-L(\mathcal S)\ge 0$ always holds.  Thus, $\Delta(\mathcal{S}) \leq \epsilon$ is equivalent to $L(\mathcal T)-L(\mathcal S)\leq \epsilon$, hence $L(\mathcal T)-\epsilon\leq L(\mathcal S) $.  Meanwhile, for a given trajectory $\mathcal{T}$, its length $L(\mathcal{T})$ is constant. So minimizing $\Delta(\mathcal{S})$ is equivalent to maximizing $L(\mathcal{S})$. Given these, the PDS and CDS problems are also equivalent to the following optimization problems, respectively. 

\begin{equation}
	\begin{aligned}
		\min_{\mathcal I \subseteq \{1,2,\ldots,N\}} \left|\mathcal I\right|\\
		\text{s. t. } L(\mathcal S) \ge L(\mathcal T) - \epsilon
		\end{aligned}
\end{equation}
}
\vspace{0.5cm}
\begin{equation}
	\begin{aligned}
	\max_{\mathcal I \subseteq \{1,2,\ldots,N\}} L(\mathcal S)\\
	\text{s.t.} \quad \left| \mathcal I \right| = k \leq K.
	\end{aligned}
\end{equation}
These alternative forms are more intuitive and straightforward to be analyzed and solved, as will be seen in the following algorithm part.

It should be mentioned that both PDS and CDS are  constrained discrete optimization problems. Solving naively involves enumerating all possible point combinations to identify the simplified trajectory of maximal length that satisfies the constraints with a minimal number of points. However, this approach has a time complexity of $O(2^N)$, which becomes computationally prohibitive as $N$ increases.

\subsection{A Unified Dynamic Programming Optimization Framework}
In this section, we prove theoretically that the above optimization problems can be solved efficiently with dynamic programming, and then propose a unified framework to solve them conveniently with one algorithm.

\begin{figure}[h]
	\centering
	\includegraphics[width=0.98\textwidth]{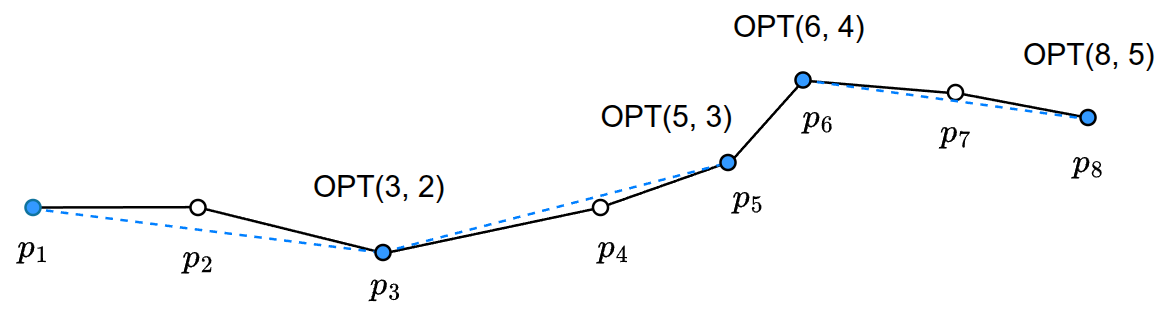}
	\caption{Optimal substructure decomposition. Blue dots and dash lines constitute the simplified trajectory.}\label{opt_substructure}
\end{figure}

It has been shown that if an optimization problem has \textbf{optimal substructure} and \textbf{overlapping subproblems} properties, then it can be solved efficiently with dynamic programming \cite{lin2019optimizing,cormen2022introduction}. More specific, a problem has optimal substructure if an optimal solution to the problem contains within it optimal solutions to its subproblems.This is the basis of our optimization algorithm. We decompose it into subproblems as follows and then prove theoretically that the CDS and PDS optimization problems have the two desired properties under such a decomposition strategy. Let $\text{OPT}(i, k)$ = \textbf{maximum total length} achievable by selecting  $k$  points from the sub-trajectory  $p_1, p_2, \dots, p_i$ , with the constraints that: {1}) the first point is  $p_0$; and the last point is  $p_i$; {2}) exactly  $k$  points are selected and the final answer to the problem is $\text{OPT}(N, K)$. An illustration is displayed in figure \ref{opt_substructure}.  We formally state it in the following theorem.

\vspace{0.3cm}
\textbf{Optimal Substructure Theorem:} \textit{Given a trajectory $\mathcal {T} = (p_1, p_2, \ldots, p_N)$,  for any  $1 < k \leq K$  and  $k \leq i \leq N$ , if  $\mathcal T' = (p_1 = p_{i_1}, p_{i_2}, \dots, p_{i_{k-1}}, p_i = p_{i_k})$  is an optimal solution for $\text{OPT}(i, k)$, then the prefix  $\mathcal T'_{\text{prefix}} = (p_1, p_{i_2}, \dots, p_{i_{k-1}})$  must be an optimal solution for $\text{OPT}(i_{k-1}, k-1)$.}
\vspace{0.3cm}

Now we prove the theorem by contradiction.

\noindent\textit{Proof:}

Assume, for contradiction, that  $\mathcal T'$  is optimal for $\text{OPT}(i, k)$, but its prefix  $\mathcal T'_{\text{prefix}}$  is not optimal for $\text{OPT}(i_{k-1}, k-1)$.
Then there exists another trajectory  $\mathcal T''_{\text{prefix}} = (p_1, p_{j_2}, \dots, p_{i_{k-1}})$  (with the same endpoint  $p_{i_{k-1}}$ ) that has greater length:
\[
L(\mathcal T''_{\text{prefix}}) > L(\mathcal T'_{\text{prefix}})
\]

Now construct a new trajectory  $\mathcal T'' = \mathcal T''_{\text{prefix}} \oplus  p_i $  (appending  $p_i$  to  $\mathcal T''_{\text{prefix}}$ ).
The length of  $\mathcal T''$  is:
\[
L(\mathcal T'') = L(\mathcal T''_{\text{prefix}}) + f(p_{i_{k-1}}, p_i)
\]
Since $L(\mathcal T''_{\text{prefix}}) > L(\mathcal T'_{\text{prefix}})$ and the last segment $f(p_{i_{k-1}}, p_i)$ is the same for both, we have:
\[
L(\mathcal T'') = L(\mathcal T''_{\text{prefix}}) + f(p_{i_{m-1}}, p_i) > L(\mathcal T'_{\text{prefix}}) + f(p_{i_{k-1}}, p_i) = L(\mathcal T')
\]
But this contradicts the assumption that  $\mathcal T'$  is optimal for $\text{OPT}(i, k)$, because we found a trajectory  $\mathcal T''$  with the same number of points ending at  $p_i$  that has strictly greater length.

Therefore, the assumption is false, and  the optimal for $\text{OPT}(i_{k-1}, k-1)$ must be $\mathcal T'_{\text{prefix}}$. $\blacksquare$

The theorem shows that the decomposition also has the overlapping property, since longer optimal simplified trajectory contains optimal shorter ones.  So, we could derive the following recurrence relation for dynamic programming to solve the CDS and PDS problems optimally. For $i > 1$ and $k \geq 1$, the optimal simplified trajectory from $p_1$ to $p_i$ with $k$  points consists of:
\begin{itemize}
	\item An optimal simplified trajectory from $p_1$ to some intermediate point $p_j$ ($1 < j < i$) with $k-1$  points.
	\item A direct segment from $p_j$ to $p_i$ that replaces all original points between $p_j$ and $p_i$.
\end{itemize}
Mathematically, let array $dp[i][k]=OPT(i, k)$, where $i\in[1,..,N], k\in[0, K]$. We have the following recurrence equations:

\begin{equation}
	\label{eq:recurrence}
	dp[i][k] = \max_{j \in \{k, k+1, \dots, i-1\}} \left( dp[j][k-1] + f(p_j, p_i) \right)
\end{equation}
with base cases:
\begin{align}
	dp[i][0] &= f(p_1, p_i) \quad \text{for all } i = 1, 2, \dots, N-1 \label{eq:base1} \\
	dp[0][k] &= 
	\begin{cases}
		0, & \text{if } k = 0 \\
		-\infty, & \text{if } k > 0
	\end{cases} \label{eq:base2}
\end{align}
The constraint $j \geq k$ in Equation~\eqref{eq:recurrence} ensures that there are at least $k$ points (including $p_j$) between $p_1$ and $p_j$ to accommodate $k-1$ points. The equation shows that we should choose the point $j$ as the next $k^{th}$ point, if it maximizes the total length of the simplified subtrajectory from $p_1$ to $p_j$ and the chord from $p_j$ to $p_i$. 

Putting all together, we propose a unified dynamic program algorithm to solve CDS and PDS shown in Algorithm \ref{UDP}. Line 4 computes the base cases (i.e. the simplified trajectory length from $p_1$ to $p_i$ with 0 interior point is the chord length between $p_1$ and $p_i$). Line 5 sets maximal number of  points in allowed the simplified trajectory. Line 8-9 build the table $dp[i][k]$ for all feasible $i$ and $k$.  The variable pred saves the corresponding optimal $j$ in equation \eqref{eq:recurrence}. Line 11-14 breaks loops if a optimal solution has been found for PDS. Line 20-25 add optimal $j$ reversely for varies number of interior points to form the final optimal simplified trajectory, i.e. an optimal simplified trajectory is always constituted of an optimal shorter simplified subtrajectory and a chord, according to the Optimal Substructure Theorem. 
\begin{algorithm}
	\caption{A Unified Dynamic Programming Algorithm}
	\label{UDP}
        \small 
	\begin{algorithmic}[1]
		\State \textbf{Input:} Trajectory $\mathcal T = \{p_i\}_{i=1}^{N}$, $\text{specified points } K$ or $\text{error tolerance }\epsilon$
		\State \textbf{Output:} Simplified trajectory indices $\mathcal I$
		
		\State \textbf{// Phase 1: DP Computation}
		\State Initialize $dp[i][0] \gets f(p_1, p_i)$ for all $i$
		\State $k_{\max} \gets$ $K$ if $K$ specified, else $N-2$ \Comment{CDS if $K$}
		
		\For{$k = 1$ to $k_{\max}$}
		\For{$i = 2$ to $N$}
		\State $dp[i][k] \gets \max_{j=k}^{i-1} \{ dp[j][k-1] + f(p_j, p_i) \}$
		\State $\text{pred}[i][k] \gets \arg\max_{j=k}^{i-1} \{ dp[j][k-1] + f(p_j, p_i) \}$
		\EndFor
		
		\If{$\epsilon$ mode and $dp[N][k] \geq L_{\text{orig}} - \epsilon$} \Comment{PDS if $\epsilon$}
		\State $k_{\text{max}} \gets k$
		\State \textbf{break}
		\EndIf
		\EndFor

		\State \textbf{// Phase 2: Backtracking}
		\State $i \gets N$
		\State $k \gets k_{\text{max}}$ \Comment{total number of points}
		\State $\mathcal I \gets [N]$
		\While{$k > 0$}
		\State $j \gets \text{pred}[i][k]$
		\State $\mathcal I \gets [j] \oplus \mathcal I$  \Comment{prepend j}
		\State $i \gets j$
		\State $k \gets k - 1$
		\EndWhile
		\State $\mathcal I \gets [1] \oplus \mathcal I$  \Comment{add first point}
		
		\State \textbf{return} $\mathcal I$
	\end{algorithmic}
\end{algorithm}

{ The global optimality of the simplified trajectory can be derived as follows. Assume that the algorithm returns the globally maximum possible length $L^* = dp[N][K]$. Suppose, to the contrary, that there exists another simplified trajectory $\mathcal{S}'$ with exactly $K$ points whose length satisfies $L(\mathcal{S}') > L^*$. 
Let $p_i$ be the penultimate point of $\mathcal{S}'$. By Optimal Substructure Theorem, the length of its prefix cannot exceed the optimal value for that prefix state, i.e., $L(\mathcal{S}'_{\text{prefix}}) \leq dp[t][K-1]$. Therefore, we have:
\[
L(\mathcal{S}') = L(\mathcal{S}'_{\text{prefix}}) + f(p_i, p_N) \leq dp[i][K-1] + f(p_i, p_N).
\]
Taking the maximum over all possible penultimate points $j$ (from $K$ to $N-1$), we get:
\[
L(\mathcal{S}') \leq \max_{j} \{ dp[j][K-1] + f(p_j, p_N) \} = dp[N][K] = L^*.
\]
This directly contradicts the assumption that $L(\mathcal{S}') > L^*$. Thus, no longer simplified trajectory exists with the same number of points, and the solution is globally optimal. 

{
It is worth mentioning that although our approach is not the first to apply dynamic programming to trajectory simplification, it differs fundamentally from existing methods in its explicit length-preserving optimization objective function. In \cite{Hu2024}, Hu et al.'s DP-based compression minimizes the overall Synchronous Euclidean Distance (SED) between the original and simplified trajectories, a pointwise positional deviation measure, and is therefore globally optimal with respect to spatial fidelity, not length; as our case studies show (Fig. 4, Table 2), SED-optimal simplification can still discard a substantial share of sailed distance, since minimizing average positional deviation does not penalize the removal of out-and-back or looping segments that contribute little to positional error but materially to path length. Similarly, Long et al.'s \cite{Long2014} direction-preserving DP framework guarantees global optimality with respect to a bearing/direction-based error, which likewise does not bound length loss. More broadly, classical error-bounded simplification methods (e.g., RDP/ARDP with a perpendicular-distance threshold, or streaming methods such as OPW and SQUISH-E) either use a local, greedy error criterion that offers no global optimality guarantee at all, or, where a global error bound is enforced, the bound is defined on positional/perpendicular deviation rather than on the cumulative length difference. In contrast, our CDS and PDS formulations take $\Delta(\mathcal I)$, the trajectory length difference itself, as the objective (CDS) or the constraint (PDS), and the dynamic programming recurrence in equations (\ref{eq:recurrence})-(\ref{eq:base2}) is constructed specifically to accumulate and maximize retained chord length rather than to minimize positional or directional deviation. The Optimal Substructure Theorem and the subsequent global-optimality analysis therefore establish optimality with respect to length preservation specifically. }

\subsection{Implementation and Parameter Tuning}

While implementing the algorithm, the major computational bottle-neck is the evaluation of the distance function $f(p_j, p_i)$ in Line 8-9, because same distance can be evaluated repeatedly. To speed-up the algorithm, for small trajectory (e.g. $N<10000$ ), we directly precompute the pairwise distance matrix $D[j,i]= f(p_j, p_i)$ for $i,j=1..N$ and then retrieve directly the distance values in line 8-9. For long trajectory, the pairwise distance matrix $D$ could be very huge, so it is not feasible to compute the whole matrix. Noticing that only a subset of pairwise distance $p_j, p_i$ need to be evaluated, so we just precompute the needed pair of distance $f(p_j, p_i)$ for $k=1..k_{max}, i=2..N$ and $j=k..i+1$. Thereafter, we directly retrieve the saved distance values in line 8-9. Other  speed-up techniques such as vectorization and matrix operation are also utilized to replace looping.
	
The parameters $K$ (target simplified length) and $\epsilon$ (target error tolerance) control the compression ratio and length error, respectively, of the simplified trajectory. They can be set to a single value for all trajectories, or, more practically, be adjusted adaptively to each trajectory based on its total data point number or total length. Based on our experiments in Section 4, there parameters are relatively robust to different navigational status or water areas. For example, from figure \ref{dataset_result} we can see that $0.1 < \epsilon < 1$ would achieve a considerable data compression rate at DCR $80-90\%$ for different loitering scenarios. Correspondingly, simplified trajectories with $80\% < K < 90\%$ of total length, the length loss rate LLR is generally less than 1\%. }

{To make the algorithm easily reproducible, we further explain the implementation in details. The DP table is initialized as $dp[0][0]=0$ and $dp[i][0]=f(p_1,p_i)$ for $i=1,\dots,N-1$ (i.e., the base case is the direct chord from the first point to $p_i$ with zero intermediate points), with all other entries set to $-\infty$ and the predecessor table $pred[i][k]$ initialized to a sentinel value ($-1$) indicating "unreached"; this sentinel is checked during backtracking to guard against infeasible states (e.g., $k > i-1$). For CDS, the recursion is evaluated for $k=1,\dots,K$ with $K$ fixed by the user, and the procedure terminates naturally once the full table has been filled, returning the trajectory reconstructed by backtracking from $dp[N-1][K]$. For PDS, the recursion is instead evaluated incrementally for $k=1,2,\dots$, and the algorithm applies an explicit early-stopping criterion: after computing row $k$, it checks whether $dp[N-1][k] \geq L(\mathcal{T})-\epsilon$ (equivalently, $\Delta(\mathcal{S}) \leq \epsilon$); the smallest $k$ satisfying this condition is taken as $k_{\max}$, the loop is terminated immediately, and the corresponding trajectory is reconstructed by backtracking — this avoids computing the full $O(N^2)$ table in the typical case. Regarding parameter settings, The algorithm only requires the target data point number $K$ or the target length error tolerance $\epsilon$.  $K$ is required to satisfy $2 \le K \le N-2$ (values outside this range are clipped) and $\epsilon > 0$ is expressed in the same units as $f$ (nautical miles in our experiments). No other tunable hyperparameters (e.g., learning rates or thresholds) are involved, since the framework is a deterministic, exact dynamic programming.}

\section{Experiments}
This section presents an experimental evaluation to compare the proposed trajectory simplification approach with state-of-the-art methods. We begin with case studies to show  the limitations of existing methods in preserving trajectory length and geometry. We then conduct experiments on a real-world trajectory dataset to further evaluate all methods.

\subsection{ Comparison Baselines and Evaluation Metrics}

Our experimental evaluation compares the proposed framework with five state-of-the-art baseline methods:
\begin{itemize}
\item The classic \textbf{Ramer-Douglas-Peucker (RDP)} algorithm\cite{Douglas1973}, valued for its computational efficiency and ability to preserve geometric shape.
\item An \textbf{Adaptive RDP (ARDP)} method \cite{tang2021method}, which improves upon standard RDP by mitigating its sensitivity to the distance threshold parameter.
\item The \textbf{Top-Down Kinematic Compression (TDKC)}  algorithm\cite{Guo2024}, developed for adaptive compression while retaining trajectory features.
\item The \textbf{Top-Down Time Ratio (TD-TR)} algorithm \cite{Meratnia2004}, which incorporates temporal characteristics into the compression process by evaluating the time-synchronous distance between the original and approximated trajectories rather than relying solely on perpendicular spatial error.
\item The global \textbf{Synchronous Euclidean Distance (SED)} minimization approach\cite{Hu2024} that  representing an optimization-based alternative.
\end{itemize}

We use the following three evaluation metrics to compare the algorithms quantitatively:
\begin{itemize}
	\item \textbf{Data Compression Ratio (DCR):} The number of points  removed from the trajectory ($N-K$) to the number of points  in the original trajectory ($N$), expressed as a percentage.
	\begin{equation}
		DCR = \frac{N - K}{N} \times 100\%
	\end{equation}
	A higher DCR value indicates a higher degree of data compression. In our experiments, we compared the DCR of different algorithms at the same length error tolerance $\epsilon$ to analyze their compression performance.
	
	\item \textbf{Length Loss Rate (LLR):} As the core metric for evaluating length preservation, LLR quantifies the percentage of the sailing distance lost in the simplified trajectory $\mathcal{S}$ compared to the original trajectory $\mathcal{T}$. It is defined as:
	\begin{equation}
		LLR = \frac{|L(\mathcal{T}) - L(\mathcal{S})|}{L(\mathcal{T})} \times 100\%
	\end{equation}
	Minimizing LLR is the primary focus of this study. In our experiments, we compare the LLR values of different algorithms under the condition of having the same number of retained points $K$.
	{
	\item \textbf{Shape Fidelity Index (SFI):}  Traditional shape similarity metrics such as the Fréchet and Hausdorff distances are not well suited to our setting for two reasons. First, the points of the simplified trajectory are strictly a subset of those of the original trajectory, so the two are not independent curves being compared but a curve and its own reduction. Second, and more importantly, both metrics are defined as the supremum (maximum) deviation over the entire curve. This worst-case nature masks the overall geometric fidelity and provides little insight into how well the simplified curve captures the original's directional structure, as we demonstrate in our experiments. Instead, inspired by the curvature-based measure proposed in \cite{gan2026sketchsemantickeypointconditioning}, we propose an integral-based metric that accumulates the local heading discrepancy along the entire path, thereby avoiding this "worst-case domination" issue. We refer to this measure as the Shape Fidelity Index (SFI), defined as:
	\begin{equation}
		SFI = \frac{K}{L(\mathcal{T})} \int_{0}^{L(\mathcal{T})} \big(1 - \cos(\theta_s - \theta'_s)\big) \, ds
	\end{equation}
	where $\theta_s$ and $\theta'_s$ denote the heading angles (tangential directions) of the original subtrajectory and the correspondingly simplified subtrajectory, respectively, parameterized by arc $s$. The factor $K/L(\mathcal{T})$ penalizes a simplified trajectory that requires more interior points to achieve the same cumulative heading discrepancy. By construction, $SFI \geq 0$, with equality if and only if $\theta_s = \theta'_s$ for every arc-length position $s$. Because SFI integrates directional deviation over the full trajectory rather than taking its maximum, it effectively penalizes algorithms that improperly smooth out critical local features such as sharp turns or small loops. A lower SFI value therefore indicates a higher degree of shape and directional consistency between the original and simplified trajectories.
	
	}
\end{itemize}

\subsection{Data Preprocessing}
{ Raw AIS data frequently contains outliers and missing values induced by network fluctuations, transmission congestion, or signal interference. Such anomalies may cause algorithms to misidentify vessel navigation states, thereby compromising analytical accuracy \cite{xie2025ais}. Consequently, we preprocessed the raw data prior to the experiments, incorporating outlier removal, missing value imputation, mooring trajectory segmentation, and chronological sorting of data points. Following the same preprocessing pipeline in \cite{xie2025ais}, the resulting point sequences represent coherent and continuous vessel voyages, which serve as the reference trajectories for subsequent comparative evaluations.}

\subsection{Case Studies}
\textbf{A. The trajectories data.} We first evaluate our algorithm through two case studies: one contains short-term vessel activities (e.g., collision avoidance) and the other contains long-term activities (e.g., loitering). Our methods achieve significantly smaller LLR for a given number of data points $K$, and greater DCR for a given error tolerance $\epsilon$.

\begin{figure}[h]
\centering
\begin{subfigure}{0.45\textwidth}
\includegraphics[width=\linewidth]{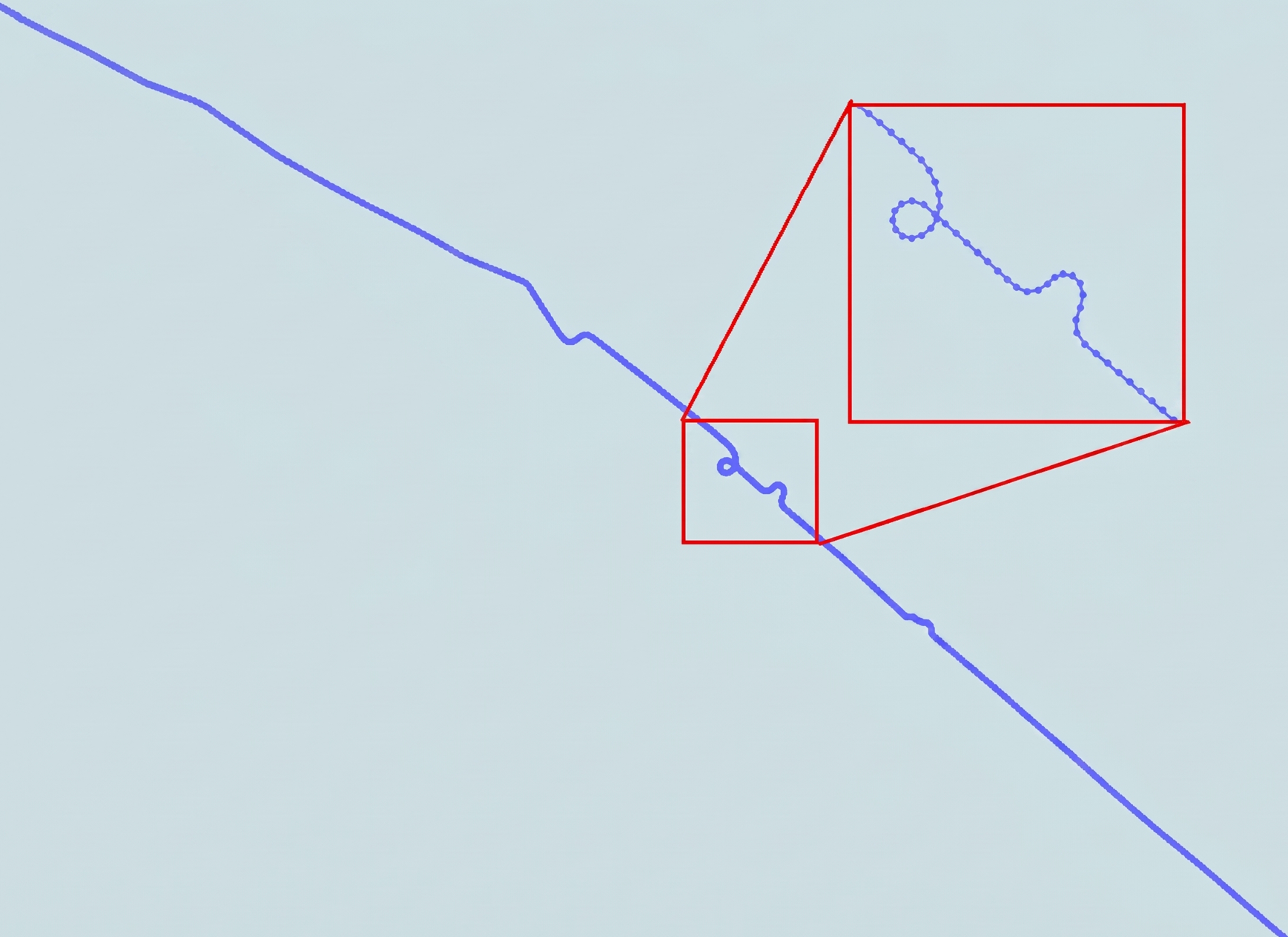}
\caption{Case trajectory 1}
\end{subfigure}
\hfill
\begin{subfigure}{0.5\textwidth}
\includegraphics[width=\linewidth]{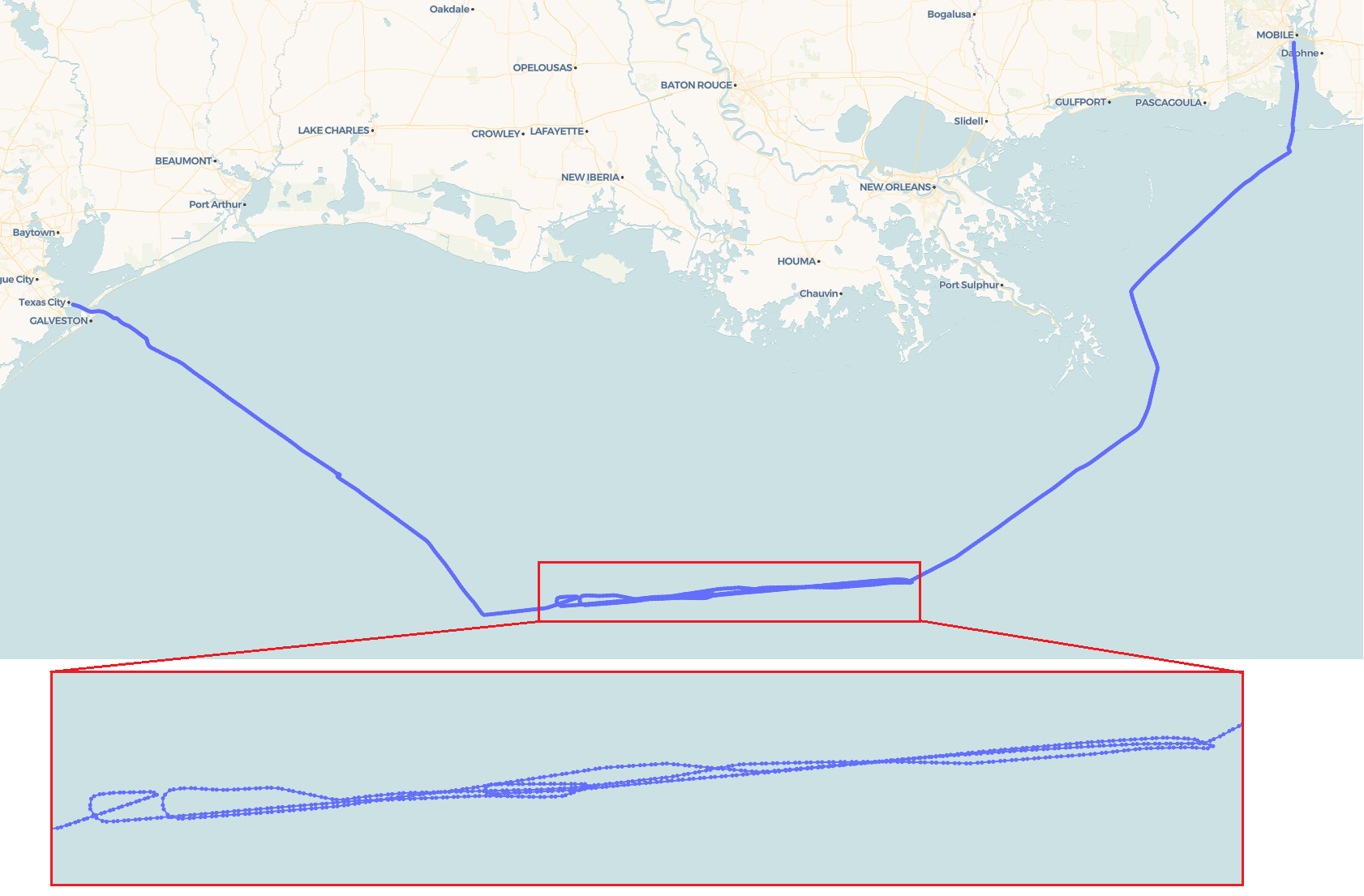}
\caption{Case trajectory 2}
\end{subfigure}

\caption{The case study trajectories}
\label{case_traj}
\end{figure}

The first case study examines a vessel trajectory characterized by predominantly straight sailing with minor local deviations. Such irregularities are frequently caused by maneuvers for collision avoidance while vessels encounter or navigation around obstacles. The trajectory's properties are listed in Table \ref{traj_info} top part, and its path plot is shown in figure \ref{case_traj} (a), in which several small, irregular local movements are visible, as highlighted and magnified within the red boxes.

\begin{table}[h]
	\centering
        \footnotesize
	\begin{tabular}{l l l l}
		\toprule
		Field & Value & \quad Field & Value\\ \hline
		MMSI & XXX810XXX & \quad Start Time & 6/15/2022 20:19:41   \\ 
		Vessel Type & Tanker & \quad End Time & 6/16/2022 12:32:11\\
		Total Points & 390 & \quad Start Location & (26.4118, -94.8372)  \\
		Total Distance (nmi) & 146.25 & \quad  End Location & (27.7899, -96.917) \\ \hline
            MMSI & XXX006XXX & \quad Start Time &   6/15/2021 23:32:00   \\ 
		Vessel Type & Cargo & \quad End Time & 6/18/2021 12:44:30\\
		Total Points & 1470 & \quad Start Location & (30.6462, -88.0319)  \\
		Total Distance (nmi) & 735.73 & \quad End Location & (29.3770, -94.8648) \\
		\bottomrule
	\end{tabular}
	\caption{Information of the case study trajectories}\label{traj_info}
\end{table}

The second case study is about vessel loitering activities, which is a common and important vessel behavior caused by scenarios such as awaiting orders,  illicit activity or avoiding potential risks \cite{zhang2022loitering, tao2025analysis, xie2025ais}. Table \ref{traj_info} bottom part provides the trajectory's information, and its overall shape is displayed in figure \ref{case_traj} (b). Note that the middle segment of the voyage shows extensive loitering, where the vessel wanders for a significant duration and distance, as highlighted and magnified within the red boxes.

\textbf{B. Experiment 1.} { The first experiment compares the proposed approach against baseline approaches using input parameters chosen by two adaptive algorithms, ARDP and TDKC. They automatically determine the optimal number of points $K$ or the error tolerance $\epsilon$ based on trajectory kinematics. For the adaptive algorithms (ARDP, TDKC), we run each algorithm to simplify the trajectories and compute the resulting precision quantities ($\epsilon$, LLR) and compression quantities ($K$, DCR). For the non-adaptive algorithms, we instead use binary search to select the input parameter such that the algorithm's output matches the compression quantities produced by the adaptive method (or the precision quantities, in the case of PDS), and then compute the corresponding precision quantities (or compression quantities, for PDS). Dashes ($-$) in the table denote the quantities that were fixed by this matching procedure rather than independently computed. The corresponding results are presented in Table \ref{tab_3}. 

\begin{table}[h!]
	\centering
	\footnotesize
	
	\caption{Results of Comparison using Automatically Chosen Parameters}\label{tab_3}
	\begin{tabular}{ccccccc}
		\toprule
		Study Cases & Method  & $\epsilon$(nmi) & LLR(\%) & $K$ & DCR(\%) & SFI \\
		\midrule
		\multirow{12}{*}{Case 1}
		& TDKC  &  0.3013 & 0.2060 & 65 & 83.33  & 0.1340 \\
		& RDP   & 0.0487 & 0.0333 & - & - & 0.0220 \\
		& SED   & 0.3098 & 0.2119 & - & - & 0.1377 \\
		& TD-TR & 0.1055 & 0.0721 & - & - & 0.0469 \\
		& CDS   & \textbf{0.0168} & \textbf{0.0115} & - & - & \textbf{0.0075} \\
		& PDS   & - & - & \textbf{31} & \textbf{92.05}  & 0.0615 \\ \cmidrule{2-7}
		& ARDP  & 0.0050 & 0.0034 & 101 & 74.10  & 0.0035 \\
		& RDP   & 0.0043 & 0.0029 & - & - & 0.0030 \\
		& SED   & 0.0461 & 0.0315 & - & - & 0.0318 \\
		& TD-TR & 0.0130 & 0.0089 & - & - & 0.0090 \\
		& CDS   & \textbf{0.0028} & \textbf{0.0019} & - & - & \textbf{0.0020} \\
		& PDS   & - & - & \textbf{87} & \textbf{77.69}  & 0.0030 \\ 
		\midrule
		\multirow{12}{*}{Case 2} 
		& TDKC  & 0.5182 & 0.0704  & 204 & 86.12 & 0.1436 \\
		& RDP   & 0.0789 & 0.0107 & - & - & 0.0219 \\
		& SED   & 0.1676 & 0.0228 & - & - & 0.0465 \\
		& TD-TR & 0.3051 & 0.0415 & - & - & 0.0846 \\
		& CDS   & \textbf{0.0342} & \textbf{0.0046} & - & - & \textbf{0.0096} \\
		& PDS   & - & - & \textbf{76}  & \textbf{94.83} & 0.0529 \\ \cmidrule{2-7}
		& ARDP  & 0.1840 & 0.0250 & 140 & 90.48  & 0.0352 \\
		& RDP   & 0.2918 & 0.0397 & - & - & 0.0557 \\
		& SED   & 0.4987 & 0.0678 & - & - & 0.0950 \\
		& TD-TR & 0.4307 & 0.0585 & - & - & 0.0821 \\
		& CDS   & \textbf{0.1070} & \textbf{0.0145} &- & - & \textbf{0.0205} \\
		& PDS   & - & - & \textbf{116} & \textbf{92.11}  & 0.0288 \\ 
		\bottomrule
	\end{tabular}
\end{table}

For Case 1, TDKC automatically selects $K=65$ points (equivalently, DCR $=83.33\%$), yielding a length error of $\epsilon=0.3013$ nautical mile (nmi) (equivalently, LLR $=0.2060\%$). We therefore fix CDS and other baseline approaches to the same value $K=65$ to directly compare its resulting precision quantities ($\epsilon$, LLR) and geometric fidelity (SFI). As shown in the table, at the same number of retained points ($K=65$, DCR $=83.33\%$), CDS achieves a substantially lower precision quantities than  that of all baseline algorithms. Concurrently, CDS attains an SFI of 0.0075, more than an order of magnitude lower than TDKC's 0.1340, and also much lower than that of other baseline algorithms. This indicates that for the same point budget, CDS preserves both trajectory length and geometric shape considerably better. A complementary comparison under PDS shows that, for a matched precision quantities ($\epsilon=0.3013$, LLR=0.2060), PDS achieves a DCR nearly 10 percentage points higher than the baseline algorithms (they share the same compression quantities with TDKC), meaning it requires substantially fewer points to approximate the original trajectory at a comparable level of length accuracy. It is worth noting that PDS has a high SFI, because its simplified trajectory using much less data points ($K=31$) to maximally compress the trajectory under the given error tolerance, and thus would certainly produce a larger shape distortion. But comparing with TDKC (same $\epsilon=0.3013$), its SFI is still much lower. The same pattern of improvement holds when comparing against ARDP, and these advantages are consistently observed for Case 2 as well.
}

\textbf{C. Experiment 2.} {The second experiment compares the proposed CDS algorithm against the baseline approaches under user-specified point-count constraints. We evaluate the resulting precision quantities ($\epsilon$, LLR) and SFI for each method as the data compression level increases, with the target DCR being 70\%, 80\%, 90\% and 95\% (respectively, the corresponding number of retained points $K$ being 30\%, 20\%, 10\%, and 5\% of the original trajectory length $N$). Results are presented in Table \ref{tab_4}, where subscripts denote the case study ID. Because TDKC and ARDP select their parameters automatically and deterministically, their simplified trajectories are identical to those obtained in Experiment 1; we therefore reproduce their results from Table \ref{tab_3} here for ease of comparison, with the two numbers preceding each entry indicating the resulting DCR for Case 1 and Case 2, respectively.

In both case studies, the LLR and SFI of the baseline algorithms increase sharply as fewer points are retained. For instance, at $K = 5\%$ in Case 1, the LLR of RDP and TD-TR approaches approximately $1\%$, while SED exceeds $1.66\%$; by contrast, CDS achieves an LLR of $0.4322\%$ — less than half that of RDP and TD-TR, and roughly one-quarter that of SED. The SFI metric reveals a consistent trend in geometric fidelity: at $K = 5\%$ in Case 1, CDS attains an SFI of 0.0819, again less than half that of RDP and TD-TR and nearly one-quarter that of SED. The results also show that SFI decreases quickly as $K$ increases. The performance advantage of CDS is even more pronounced in Case 2, where CDS consistently achieves the lowest error across all compression levels. We note that, because the two adaptive methods retain substantially more points than the fixed-$K$ settings evaluated here, their error metrics are not directly comparable to those of the other approaches and are included for reference only.}

\begin{table}[h!]
	\centering
	\footnotesize
	
	\caption{Results of specifying number of data points}\label{tab_4}
	\begin{tabular}{cccccccc}
		\toprule
		DCR & Method  & $\epsilon_1$(nmi) & LLR$_1$(\%) & SFI$_1$ & $\epsilon_2$(nmi) & LLR$_2$(\%) & SFI$_2$ \\ 
		\midrule
		\multirow{4}{*}{70}
		& RDP  & 0.0027 & 0.0018 & 0.0022 & 0.0080 & 0.0011 & 0.0048 \\
		& SED  & 0.0173 & 0.0118 & 0.0138 & 0.1277 & 0.0174 & 0.0766 \\
		& TD-TR  & 0.0094 & 0.0064 & 0.0075 & 0.2356 & 0.0320 & 0.1412 \\
		& CDS & \textbf{0.0019} & \textbf{0.0013} & \textbf{0.0015} & \textbf{0.0033} & \textbf{0.0005} & \textbf{0.0020} \\ 
		\cmidrule{1-8}
		\multirow{4}{*}{80} 
		& RDP & 0.0164 & 0.0112 & 0.0088 & 0.0281 & 0.0038 & 0.0113 \\
		& SED  & 0.1606 & 0.1098 & 0.0856 & 0.1330 & 0.0181 & 0.0532 \\
		& TD-TR  & 0.0411 & 0.0281 & 0.0219 & 0.2896 & 0.0394 & 0.1154 \\
		& CDS  & \textbf{0.0082} & \textbf{0.0056} & \textbf{0.0044} & \textbf{0.0116} & \textbf{0.0016} & \textbf{0.0047} \\ 
		\cmidrule{1-8}
		\multirow{4}{*}{90}         
		& RDP  & 0.3490 & 0.2386 & 0.0931 & 0.2754 & 0.0374 & 0.0556 \\
		& SED  & 0.6548 & 0.4477 & 0.1746 & 0.4628 & 0.0629 & 0.0925 \\
		& TD-TR  & 0.4907 & 0.3355 & 0.1309 &  0.4125 & 0.0561 & 0.0826 \\
		& CDS &  \textbf{0.1485} & \textbf{0.1015} & \textbf{0.0396} & \textbf{0.0939} & \textbf{0.0128} & \textbf{0.0190} \\ 
		\cmidrule{1-8}
		\multirow{4}{*}{95} 
		& RDP  & 1.3995 & 0.9569 & 0.1817 & 2.8790 & 0.3913 & 0.2858 \\
		& SED  & 2.4278 & 1.6601 & 0.3154 & 2.9525 & 0.4013 & 0.2930 \\
		& TD-TR  & 1.3730 & 0.9389 & 0.1783 & 1.4431 & 0.1962 & 0.1434 \\
		& CDS  & \textbf{0.6321} & \textbf{0.4322} & \textbf{0.0819} & \textbf{0.5586} & \textbf{0.0759} & \textbf{0.0559} \\ 
		\midrule
		83.33|86.12 & TDKC  & 0.3013 & 0.2060 & 0.1340 & 0.5182 & 0.0704 & 0.1436 \\
		74.10|90.48 & ARDP  & 0.0050 & 0.0034 & 0.0035 & 0.1840 & 0.0250 & 0.0352 \\
		\bottomrule
	\end{tabular}
\end{table}

\begin{figure}[h]
	\centering%
	\includegraphics[width=0.98\textwidth]{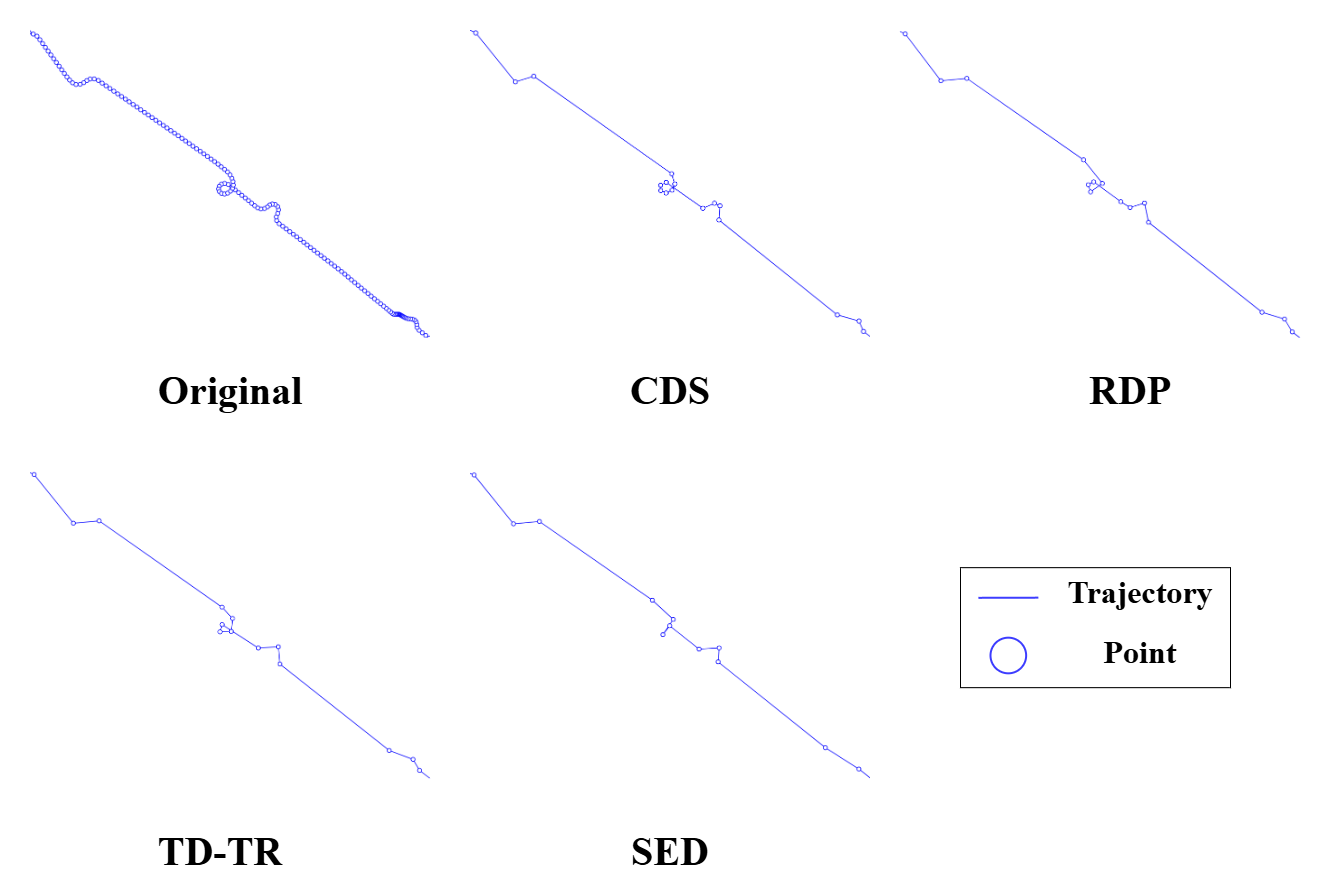}
	\caption{Simplified trajectories generated by different algorithms for case study 1.}\label{fig_5}
\end{figure}

To illustrate the qualitative differences among the methods, figure \ref{fig_5} presents the simplified trajectories obtained when retaining only \(k = 5\%\) of the original points in case 1 \footnote{The figure only shows the middle part of the trajectory for clear visualization. We put the full simplified trajectory of case 1 and case 2 in the appendix due to space limitation.}. As can be observed, SED fails to preserve the loop structure entirely. Although RDP and TD-TR maintain the loop, their geometric fidelity remains substantially compromised, with noticeable distortions in shape. In contrast, only CDS successfully preserves the loop. {The SFI of CDS is also significantly lower than that of the baselines. 
As mentioned earlier, Fréchet distance is a single bottleneck (max) statistic over the entire trajectory. Take case 1 as an example, the bottleneck for both RDP and CDS sits near the very first long chord — because that first straight segment spans ~1/3 of the whole track with zero interior support after simplification (see appendix for full simplified trajectory). That single early deviation is an order of magnitude larger than anything the loop produces. So the loop is invisible to Fréchet distance simply because it was never the worst offender — Fréchet has no mechanism to ``notice'' a well-preserved local feature buried under a much larger global deviation, even through CDS preserves the loop structure much better.}


\textbf{D. Experiment 3.} To evaluate the efficacy of the proposed PDS algorithm under stringent compression constraints, a third experiment was conducted comparing its performance against the baseline methods across prescribed length error tolerances \(\epsilon \in \{2.0, 1.0, 0.5, 0.1\}\) nautical miles. A fundamental trade-off exists between compression ratio and geometric fidelity, wherein a greater reduction in the number of retained points inevitably incurs a larger length deviation. Consequently, under a fixed error bound, the relative superiority of a given method is directly indicated by its capacity to achieve maximal data reduction. 

The quantitative results, detailed in Table \ref{tab_merge}, demonstrate that PDS consistently surpasses all baseline algorithms across the entire spectrum of tolerance levels, with the margin of improvement becoming markedly more substantial as the permissible error diminishes. This trend is particularly evident in Case 2 at the strictest tolerance of \(\epsilon = 0.1\) nmi. Under this condition, the TD-TR algorithm produces very low compression, retaining \(K = 1262\) points and yielding a DCR of merely \(14.15\%\), whereas SED achieves a DCR of \(62.24\%\). In contrast, PDS satisfies the same rigorous \(0.1\) nmi bound while retaining only \(K = 144\) points, corresponding to a significantly higher DCR of \(90.20\%\). For the same level of length error $\epsilon$, the SFI of PDS is also much lower than that of baseline approaches. The results underscore the substantial advantage of PDS in preserving trajectory integrity while maintaining significant compression rates, particularly in low-tolerance regimes. {Similar to Experiment 2,  because the two adaptive methods use self-determined $\epsilon$ to retain more points than other approaches, their error metrics are not directly comparable to those of the other approaches and are included for reference only.}

\begin{table}[h!]
	\centering
	\footnotesize
	
	\caption{ Results of specifying length error tolerance}\label{tab_merge}
	\begin{tabular}{cccccccc}
		\toprule
		$\epsilon$ (nmi) & Method & $K_1$ & DCR$_1$ (\%) & SFI$_1$ & $K_2$ & DCR$_2$ (\%) & SFI$_2$ \\
		\midrule
		\multirow{4}{*}{2.0} 
		& RDP & 14 & 96.41 & 0.1775 & 77 & 94.76 & 0.1951 \\
		& SED & 21 & 94.62 & 0.2767 & 83 & 94.35 & 0.2162 \\
		& TD-TR & 15 & 96.15 & 0.1891 & 65 & 95.58 & 0.1760 \\
		& PDS & \textbf{11} & \textbf{97.18} & \textbf{0.1248} & \textbf{37} & \textbf{97.48} & \textbf{0.1014} \\ 
		\cmidrule{1-8}  
		\multirow{4}{*}{1.0} 
		& RDP & 24 & 93.85 & 0.1396 & 86 & 94.15 & 0.1124 \\
		& SED & 36 & 90.77 & 0.2291 & 118 & 91.97 & 0.1508 \\
		& TD-TR & 29 & 92.56 & 0.1826 & 94 & 93.61 & 0.1261 \\
		& PDS & \textbf{15} & \textbf{96.15} & \textbf{0.0999} & \textbf{55} & \textbf{96.26} & \textbf{0.0733} \\ 
		\cmidrule{1-8}
		\multirow{4}{*}{0.5} 
		& RDP & 33 & 91.54 & 0.1060 & 111 & 92.45 & 0.0714 \\
		& SED & 45 & 88.46 & 0.1411 & 140 & 90.48 & 0.0949 \\
		& TD-TR & 38 & 90.26 & 0.1277 & 134 & 90.88 & 0.0836 \\
		& PDS & \textbf{23} & \textbf{94.10} & \textbf{0.0766} & \textbf{77} & \textbf{94.76} & \textbf{0.0521} \\ 
		\cmidrule{1-8}
		\multirow{4}{*}{0.1} 
		& RDP & 57 & 85.38 & 0.0351 & 187 & 87.28 & 0.0251 \\
		& SED & 83 & 78.72 & 0.0534 & 555 & 62.24 & 0.0752 \\
		& TD-TR & 66 & 83.08 & 0.0394 & 1262 & 14.15 & 0.1693 \\
		& PDS & \textbf{43} & \textbf{88.97} & \textbf{0.0275} & \textbf{144} & \textbf{90.20} & \textbf{0.0196} \\ 
		\midrule
		0.3013|0.5182 & TDKC & 65 & 83.33 & 0.1340 & 204 & 86.12 & 0.1436 \\
		0.0050|0.1840 & ARDP & 101 & 74.10 & 0.0035 & 140 & 90.48 & 0.0352 \\ 
		\bottomrule
	\end{tabular}
\end{table}

\subsection{Dataset Evaluations}
\label{sec_dataset}
\begin{figure}[h!]
	\centering%
	\includegraphics[width=0.98\textwidth]{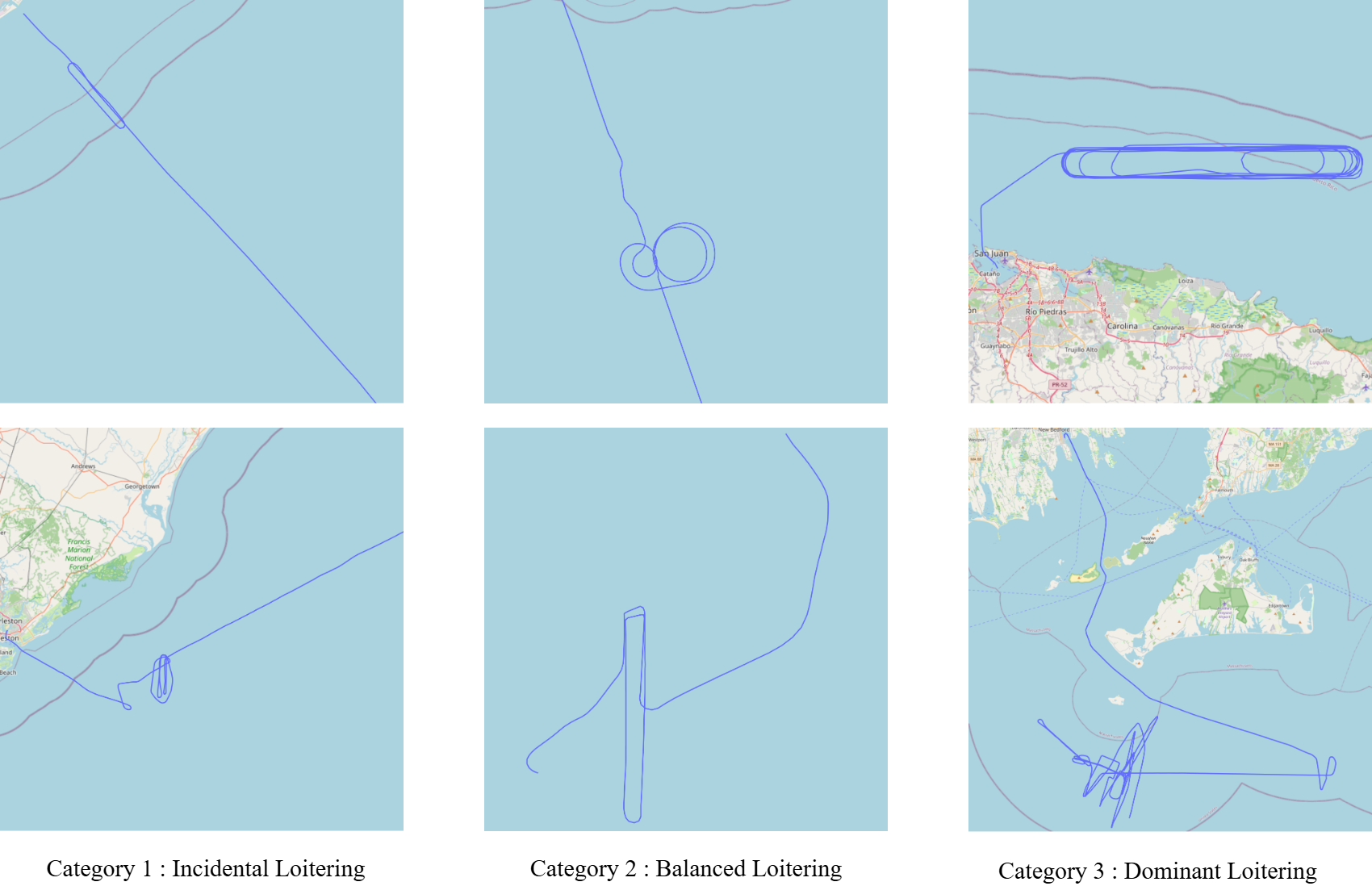}
	\caption{Sample trajectories in the dataset.}\label{fig_7}
\end{figure}

To ensure a more comprehensive evaluation, further experiments are conducted using the publicly available AIS vessel loitering dataset \cite{xie2025ais}. This dataset, characterized by highly nonlinear and spatially overlapping trajectories, presents substantial challenges for simplification. The trajectories are partitioned into three categories based on loitering intensity, spatial extent, and proportion (Fig. \ref{fig_7}). For statistical robustness, 50 trajectories are randomly sampled from each category:
\begin{itemize}
    \item \textbf{Category 1 (Incidental Loitering)}: Predominantly straight-line navigation with brief, localized loitering episodes.  
\item \textbf{Category 2 (Balanced Loitering)}: Comparable proportions of normal sailing and loitering, with loitering covering broader areas and longer durations.  
\item  \textbf{Category 3 (Dominant Loitering)}: Extensive, high‑density loitering with repetitive overlapping patterns, causing extreme data redundancy.
\end{itemize}



	
	

\begin{figure}[h]
	\centering
	\begin{subfigure}{0.98\textwidth}
		\includegraphics[width=\linewidth]{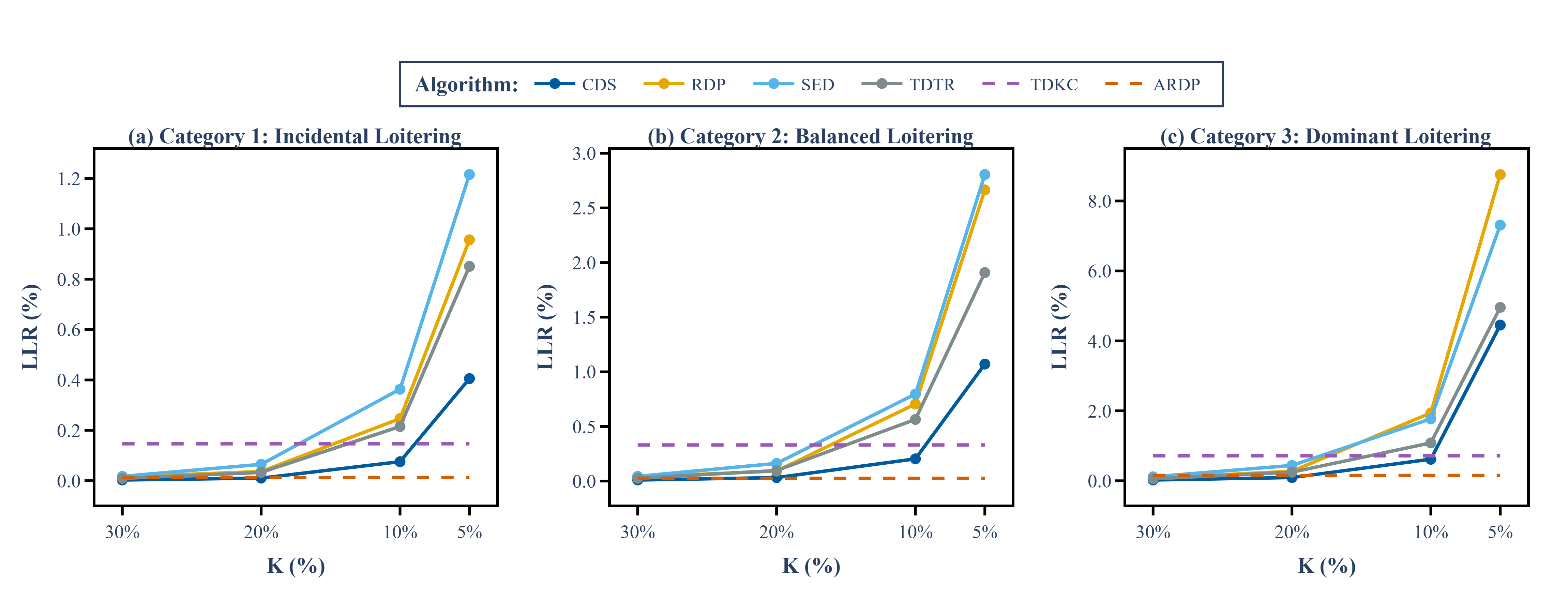}
		\caption{LLR Under Varying Retention Points $K$ ($K$ of TDKC and ARDP in Table \ref{tab_adaptive_params}).}
	\end{subfigure}
	\hfill
	\begin{subfigure}{0.98\textwidth}
		\includegraphics[width=\linewidth]{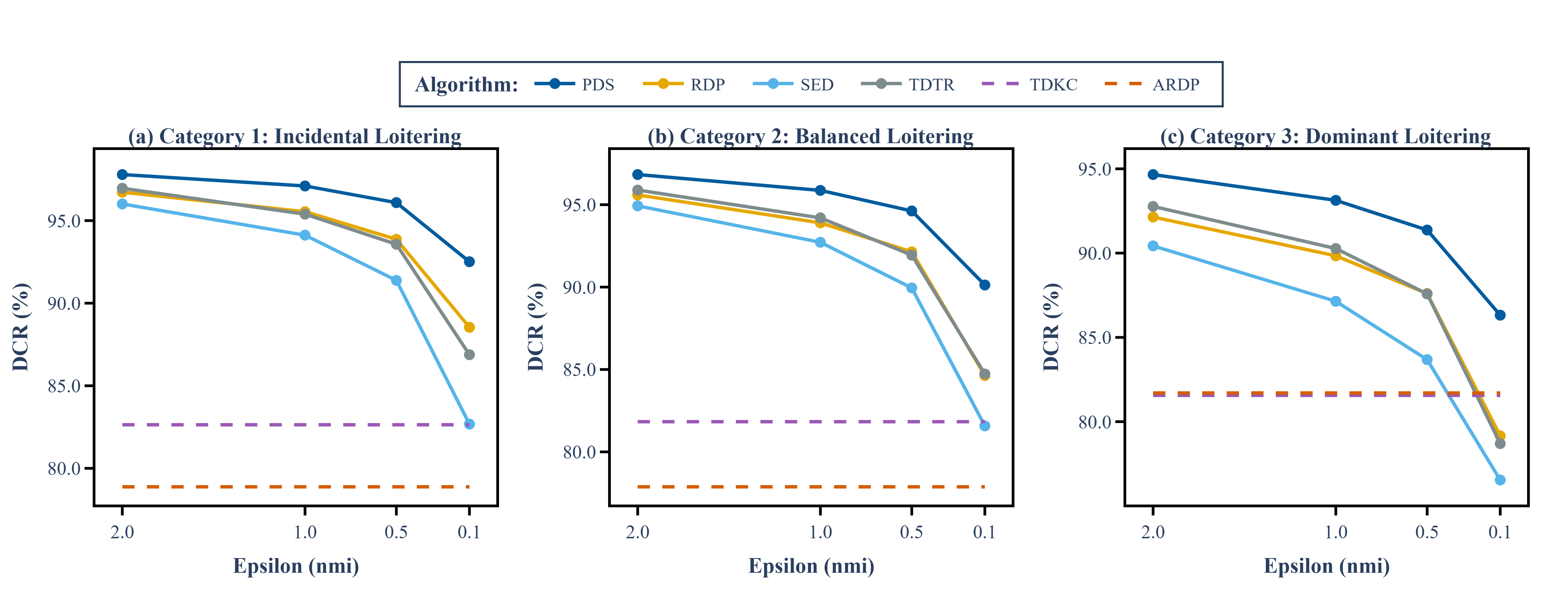}
		\caption{DCR Under Varying Error Tolerances $\epsilon$ ($\epsilon$ of TDKC and ARDP in Table \ref{tab_adaptive_params}).}
	\end{subfigure}
	
	\caption{Dataset compression results}
	\label{dataset_result}
\end{figure}

{ The results are displayed in figure \ref{dataset_result}.  Again, because the two adaptive algorithms TDKC and ARDP automatically determine their optimal point number $K$ and/or error tolerance $\epsilon$, and their results are always the same for a given trajectory, displayed as  constant flat lines in the figure, and their averaged parameters are listed in Table \ref{tab_adaptive_params}.
	
\begin{table}[h]
	\centering
	\footnotesize
	\caption{Average $K$ and $\epsilon$ of TDKC and ARDP in three categories}\label{tab_adaptive_params}
	\makebox[\textwidth][c]{
		\begin{tabular}{ccccccc}
			\toprule
			\multirow{2}{*}{Parameter} & \multicolumn{2}{c}{Category 1} & \multicolumn{2}{c}{Category 2} & \multicolumn{2}{c}{Category 3} \\
			& TDKC & ARDP & TDKC & ARDP & TDKC & ARDP\\
			\midrule
			$K(\%)$ & 17.4 & 21.12 & 18.14 & 22.12 & 18.44 & 18.30 \\
			$\epsilon$ & 0.1825 & 0.0135 & 0.3614 & 0.0323 & 1.4764 & 0.4185 \\
			\bottomrule
		\end{tabular}
	}
\end{table}
}
%
	
 Figure (a) reveals the vulnerability of the baseline algorithms in scenarios involving extreme data reduction. As the allocated number of retained points $K$ decreases from 30\% to 5\% of the original trajectory size, the length loss rate of all algorithms naturally increases; however, the CDS algorithm maintains the most stable error growth curve. Under the extremely restrictive budget where only 5\% of the points are retained, the performance gap becomes striking, particularly for category 3, where massive overlapping trajectories are prevalent. In this context, distance-independent methods  suffer severe length distortions, especially RDP whose LLR exceeds 8\%, whereas the LLR for CDS method is less than 5\%. This finding highlights the necessity of explicitly formulating the preservation of length as the primary optimization objective when processing highly dense and repetitive loitering behaviors.

In the precision-driven evaluation, figure \ref{dataset_result}(b) reveals that the proposed algorithm demonstrates superior structural retention capabilities when subjected to increasingly strict error tolerances. When the threshold of epsilon tightens from 2.0 to 0.1 nmi, the compression ratios of the baseline algorithms experience steep declines. For instance, in the Dominant Loitering scenario, relaxing the tolerance to 2.0 nmi allows all methods to achieve a compression ratio exceeding 90\%. However, restricting the tolerance to 0.1 nmi causes the compression ratio of SED to drop below 80\%. Conversely, the performance of the PDS method degrades gracefully, maintaining a compression ratio above 85\% even at the strictest bound of 0.1 nmi in Scenario 3. It consistently outperforms all competing methods, and the results demonstrate its potential applicability to high-fidelity maritime auditing tasks.

\subsection{Complex Waters Evaluations}

{
	To evaluate the generalization capability of the proposed method in complex, real-world maritime environments over a continuous time period, we collected AIS data from the San Pedro Bay port complex between August 1 and August 7, 2021. As one of the busiest and most intricate port complexes in the world, this region experiences peak maritime traffic during August. The collected dataset spans five vessel categories, including cargo vessels, tankers, tugboats, passenger ships, and fishing vessels, comprising a total of 199 vessels, 350 trajectories, and 171,965 data points.
	As a demonstration, we report results only for $K=5\%$ and $\epsilon=0.1$ nmi, rather than exhaustively varying both parameters to produce a large but repetitive table. Results are summarized in Table \ref{port_area_results}.

	\begin{table}[h!]
		\centering
		\footnotesize
		
		\caption{Experimental results on the San Pedro Bay port}\label{port_area_results}
		\makebox[\textwidth][c]{
			\begin{tabular}{ccccccc}
				\toprule
				Setting & Method & $\epsilon$ (nmi) & LLR (\%) & $K$ & DCR (\%) & SFI \\
				\midrule
				\multirow{4}{*}{$K=5\%$} 
				& RDP   & 29.1867  & 6.8949  & - & - & 1.0804 \\
				& SED   & 40.5726 & 10.0050  & - & - & 1.5130 \\
				& TD-TR & 17.6747  & 6.8064  & - & - & 0.7879 \\
				& CDS   & \textbf{16.4249}  & \textbf{5.2215}  & - & - & \textbf{0.6076} \\ 
				\midrule
				\multirow{4}{*}{$\epsilon=0.1$ nmi} 
				& RDP   & - & - & 135  & 77.79  & 0.1144 \\
				& SED   & - & - & 75  & 84.25  & 0.1034 \\
				& TD-TR & - & - & 65  & 85.99  & 0.0956 \\
				& PDS   & - & - & \textbf{36}  & \textbf{91.62}  & \textbf{0.0628} \\
				\midrule
				\multirow{2}{*}{Adaptive} 
				& TDKC  & 12.3878  & 3.0128  & 90  & 79.75  & 2.6977 \\
				& ARDP  & 1.2087  & 0.3549  & 55  & 79.98  & 0.4183 \\
				\bottomrule
			\end{tabular}
		}
	\end{table}
	
	The proposed method exhibits a clear advantage in preserving trajectory length under such complex traffic conditions. Under the fixed point-budget setting ($K=5\%$), CDS achieves the lowest mean length error, with an LLR of 5.22\%, compared with 6.89\% for RDP, 6.81\% for TD-TR, and 10.01\% for SED. Under the fixed error-tolerance setting ($\epsilon=0.1$ nmi), PDS attains the highest mean compression, with a DCR of 91.62\%, retaining an average of only 36 points per trajectory---versus 135, 65, and 75 points required by RDP, TD-TR, and SED, respectively, to satisfy the same tolerance. Beyond cumulative length preservation, the proposed framework exhibits superior geometric fidelity. As reported in Table \ref{port_area_results}, under the setting $K = 5\%$, CDS attains the lowest SFI value of 0.6076, outperforming RDP (1.0804) and SED (1.5130). Similarly, under the setting $\epsilon = 0.1$ nmi, PDS achieves the lowest SFI of 0.0628, confirming the capability of the proposed method to preserve trajectory geometry even in complex navigational environments.
	
	Because the two adaptive algorithms, TDKC and ARDP, determine their controlling parameters ($K$ or $\epsilon$) automatically rather than accepting a user-specified target, their results are reported separately for reference rather than under a matched setting. TDKC retains an average of 90 points per trajectory (DCR = 79.75\%) at a mean LLR of 3.0128\%; by comparison, CDS  attains a slightly lower length accuracy (LLR = 5.22\%) while retaining only 5\% of the original points (DCR = 95\%)---approximately one quarter as many points as TDKC for comparable length fidelity. ARDP produces a comparable number of points with PDS ($k_{ARDP}=55$ v.s. $k_{PDS}=36$), but PDS length error and SFI are much smaller than that of ARDP ($\epsilon_{PDS}=0.1$ v.s. $\epsilon_{ARDP}=1.2087$). Taken together, these results confirm that the proposed CDS/PDS framework generalizes effectively beyond the individual case studies and the loitering benchmark to a continuous, multi-vessel-type AIS dataset from a complex port environment.
	
	{
	We also conduct an experiment to further study how compression ratio affects trajectory geometric distortion. We vary $K$ from 5\% to 50\% and compare the three geometric distortion metrics: length error $\epsilon$, LLR and SFI. Results are show in figure \ref{fig_ratio_distortion}. From the results we can see that for this dataset, 80\%-85\% of compression ratio could generate simplified trajectories with considerably small geometric distortion. The results also show that the SFI has stronger capability of distinguishing small geometric distortions than $\epsilon$ and LLR. For example, the LLR curves of RDP and TD-TR are also most indistinguishable in the middle figure, but SFI shows clearly that TD-TR produce smaller accumulated heading discrepancy, hence better than RDP. Note that the proposed algorithm CDS achieves the best results across all compression ratio for all metrics.
	\begin{figure}[h]
		\centering%
		\includegraphics[width=0.98\textwidth]{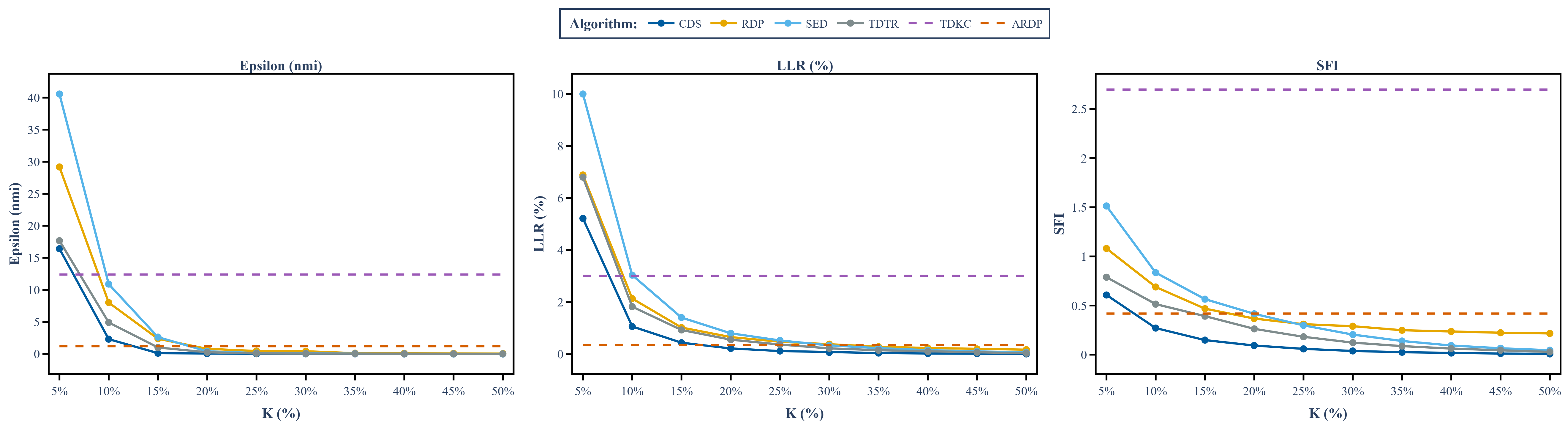}
		\caption{Compression ratio effect study}\label{fig_ratio_distortion}
	\end{figure}
}
}

\subsection{Statistical Significance Analysis}

{

To rigorously evaluate the performance advantages of the proposed framework, we conducted statistical significance testing on the 350 trajectories from the San Pedro Bay dataset. Initially, a paired-sample $t$-test was considered to compare the performance of our methods against baseline approaches. However, normality tests revealed that the paired differences significantly deviated from a Gaussian distribution ($p < 0.05$). Consequently, to ensure methodological rigor, we employed the nonparametric Wilcoxon signed-rank test with Bonferroni correction for multiple comparisons. The detailed test results are summarized in Table \ref{tab_significance}. We selected RDP and TD-TR as representative baselines because they provide the strongest competing performance under the corresponding evaluation settings. The results are sufficient to demonstrate that CDS and PDS significantly outperform these strong approaches via rigorous testing.

Under the CDS mode ($K = 5\%$), we evaluated the absolute length error $\epsilon$. As reported in Table 5, RDP attained a lower LLR, whereas TD-TR achieved a smaller $\epsilon$ among the baseline methods. We therefore benchmarked CDS against both RDP and TD-TR. The Wilcoxon signed-rank test confirmed that the proposed CDS scheme achieves significantly lower length errors than both RDP ($W = 953.0, p < 0.001$) and TD-TR ($W = 1117.0, p < 0.001$).

Under the PDS mode ($\epsilon = 0.1$ nmi), we evaluated the number of retained trajectory points $k$. Because RDP was the top-performing baseline under this configuration, we conducted a direct comparison between PDS and RDP. The statistical test revealed that PDS requires significantly fewer data points than RDP while strictly satisfying the identical error bound ($W = 0.0, p < 0.001$).

Across both modes, the resulting $p$-values demonstrate strong statistical significance ($p < 0.001$). These empirical findings confirm that the capacity of the proposed framework to concurrently maximize trajectory length preservation and minimize data volume is statistically robust rather than an artifact of random variation.

\begin{table}[t]
	\centering
	\footnotesize
	\caption{Statistical significance tests}\label{tab_significance}
	\begin{tabular}{llcccc}
		\toprule
		Mode & Metric & Baseline & W-Statistic & Adjusted $p$-value & Significance \\
		\midrule
		\multirow{2}{*}{CDS ($K=5\%$)} & \multirow{2}{*}{$\epsilon$} & RDP & 953.0 & $< 0.001$ & Yes \\
		& & TD-TR & 1117.0 & $< 0.001$ & Yes \\
		\midrule
		PDS ($\epsilon=0.1$ nmi) & $K$ & RDP & 0.0 & $< 0.001$ & Yes \\
		\bottomrule
	\end{tabular}
\end{table}

}

\subsection{Application}
Trajectory length is not merely a geometric property but a critical regulatory parameter in maritime applications. To demonstrate the practical value of length-preserving trajectory simplification, this section evaluates the influence of these algorithms on downstream applications, particularly the Energy Efficiency Operational Indicator (EEOI).
The EEOI is the primary metric recommended by the International Maritime Organization (IMO) for assessing the operational carbon intensity of individual vessel voyages. The core calculation of this metric relies on the ratio of total $CO_{2}$ emissions to the transport work performed, which is defined as the product of cargo capacity and the actual sailing distance.  

When trajectory simplification algorithms compress AIS data, they often introduce spatial shortcuts that artificially reduce the calculated actual distance. Because the actual time elapsed and the fuel consumed during the voyage remain constant, this reduction in distance inevitably leads to an inflated EEOI value. To quantify this impact, we utilized the trajectory from Case 2. Assuming that the vessel navigates in calm waters and ignoring the effects of weather and sea states, we employed the VERDE model \cite{tvete2020modelling}, which is based on hydrodynamic principles, to calculate the energy consumption of the vessel. 


Under this controlled setup, the cumulative fuel consumption was multiplied by a standard fuel conversion factor to obtain the total $CO_{2}$ emissions, which served as the constant numerator in the calculation of the EEOI. Furthermore, we constrained all algorithms to only 5\% of the original data points. The simplified trajectory lengths generated by each algorithm were subsequently substituted into the EEOI formula to calculate the corresponding assessment errors (calculation procedure in appendix). The errors can be accumulated in a year and will have significant influence upon evaluation of vessel effects upon environment and carbon reduction, such as  Carbon Intensity Indicator (CII) \cite{bayraktar2023scenario}.  The results of trajectory simplification on the calculation of the EEOI are presented in Table \ref{tab_9}. The first row is the original full trajectory without simplification. From the results we can see that both the trajectory length error and the  EEOI Error introduced by the proposed CDS are much lower than that of the other approaches.

\begin{table}[h]
	\centering
	\caption{EEOI assessment errors under $K=5\%$}\label{tab_9}
	\begin{tabular}{cccc}
		\toprule
		$K$(\%) & Method & $\epsilon$(nmi)  & EEOI Error (\%) \\ 
		\midrule
		100 & Original & 0  & 0.00 \\
		\midrule
		\multirow{4}{*}{5} 
		& RDP & 2.8790  & 0.39 \\
		& SED & 2.9525  & 0.40 \\
		& TD-TR & 1.4431  & 0.20 \\
		& CDS & \textbf{0.5586}  & \textbf{0.08} \\ 
		\bottomrule
	\end{tabular}
\end{table}
{
It is important to note that the proposed method does not itself reduce a vessel's fuel consumption or CO$_2$ emissions; rather, it reduces the bias that trajectory simplification introduces into the estimation of these quantities, which in turn supports carbon-reduction efforts indirectly but substantively. Balcer et al. \cite{Balcer2024} demonstrate this link concretely: using a bottom-up, hydrodynamics-based emissions model on 60 real ship trajectories, they show that CO$_2$ estimation error grows systematically with the degree of trajectory simplification. Since regulatory frameworks such as the IMO DCS, EU MRV, and CII rating all rely on AIS-derived distance and fuel estimates for compliance verification and carbon-intensity classification, any systematic underestimation of sailed distance directly propagates into underestimated reported emissions and can mask genuinely inefficient voyages from the auditing process. By minimizing the length-approximation error at a given compression ratio (or minimizing the number of retained points at a given length-error tolerance), our CDS/PDS framework could make contributions to reduce this source of assessment bias specifically, thus improving the fidelity of the auditing and decision-support layer of maritime energy management.

In addition, the EEOI evaluation here is conducted under calm-water assumption and does not explicitly consider environmental factors such as wind, waves, and ocean currents, which may affect vessel speed, fuel consumption, and emissions. Therefore, the results should be interpreted as quantifying the impact of trajectory simplification on EEOI estimation under controlled conditions, rather than as a comprehensive assessment of real-world fuel consumption or emissions. Future work will incorporate environmental and operational factors to evaluate the robustness of the proposed methods under more realistic maritime conditions.

\subsection{Computational Complexity Analysis}
In this section, we analyze the theoretical computational complexity and perform an empirical comparison of its execution time against baseline methods.
The computational complexity is primarily determined by the total number of original trajectory points $N$ and the number of retained points $K$.

\begin{itemize}
	
\item \textbf{CDS mode}: The algorithm maintains a DP table of dimensions $N \times K$, yielding a space complexity of $O(N \cdot K)$. The overall time complexity is governed by three nested loops: the outer loop iterates $K$ times, the middle loop traverses $N$ points, and the innermost loop searches for the optimal predecessor in at most $O(N)$ operations. Consequently, this leads to a worst-case time complexity of $O(K \cdot N^2)$. Since $K \ll N$ in typical trajectory simplification scenarios, the practical computational cost remains well constrained.

\item \textbf{PDS mode}: Since the final number of retained points is not predetermined, the maximum size of the DP table reaches $N \times N$, which results in a space complexity of $O(N^2)$. Theoretically, the worst-case time complexity is $O(N^3)$. To mitigate this overhead, we introduce an early stopping mechanism: once the simplified trajectory satisfies the predefined length error tolerance $\epsilon$, the algorithm terminates execution immediately. This strategy significantly suppresses redundant state evaluations and effectively bounds the actual execution time.

\end{itemize}

To evaluate practical efficiency, we measured the average execution time per trajectory on the dataset in Section \ref{sec_dataset}. Table \ref{tab_rt} summarizes the empirical execution times of all evaluated algorithms. The three parts of the table in corresponds to runtime on Category 1, 2 and 3 of the dataset.
\begin{table}[h]
	\centering
	\footnotesize
	\caption{Experimental results of execution time (s) of category 1-3 on the loitering dataset.}\label{tab_rt}
	\makebox[\textwidth][c]{
		\begin{tabular}{ccccccccc}
			\toprule
			 \multirow{2}{*}{Method} & \multicolumn{4}{c}{$K$ (\%)} & \multicolumn{4}{c}{$\epsilon$ (nmi)} \\
			\cmidrule(lr){2-5} \cmidrule(lr){6-9}
			& 30 & 20 & 10 & 5 & 2.0 & 1.0 & 0.5 & 0.1 \\
			\midrule
			 CDS   & 0.1751 & 0.1307 & 0.0678 & 0.0377 & - & - & - & -  \\
			 PDS   & - & - & - & - & 0.3799 & 0.3740 & 0.3559 & 0.3471  \\
			RDP   & 0.2893 & 0.2770 & 0.2168 & 0.1687 & 0.2977 & 0.3039 & 0.3046 & 0.3269  \\
			SED   & 0.9337 & 0.9129 & 0.7872 & 0.7465 & 8.5970 & 8.4839 & 8.4411 & 8.8355  \\
			TD-TR & \textbf{0.0215} & \textbf{0.0195} & \textbf{0.0131} & \textbf{0.0095} & \textbf{0.0176} & \textbf{0.0200} & \textbf{0.0209} & \textbf{0.0238}  \\
			\cmidrule(lr){2-9}
			TDKC  &\multicolumn{8}{c}{0.0185}   \\
			ARDP  &\multicolumn{8}{c}{0.0016} \\
			\midrule
			CDS   & 0.1946 & 0.1643 & 0.0794 & 0.0443 & - & - & - & - \\
			PDS   & - & - & - & - & 0.4014 & 0.4192 & 0.3780 & 0.4122 \\
		    RDP   & 0.3258 & 0.2613 & 0.2043 & 0.1583 & 0.2881 & 0.2967 & 0.2929 & 0.3355 \\
			SED   & 0.8990 & 0.9457 & 0.7740 & 0.7369 & 7.9806 & 8.1214 & 7.8117 & 9.0474 \\
			TD-TR & \textbf{0.0211} & \textbf{0.0204} & \textbf{0.0132} & \textbf{0.0096} & \textbf{0.0177} & \textbf{0.0200} & \textbf{0.0202} & \textbf{0.0243} \\
			\cmidrule(lr){2-9}
			TDKC  & \multicolumn{8}{c}{0.0166} \\
			ARDP  & \multicolumn{8}{c}{0.0015} \\
			\midrule
			CDS   & 0.9064 & 0.6682 & 0.3784 & 0.2213 & - & - & - & - \\
			PDS   & - & - & - & - & 1.7549 & 1.8282 & 1.7980 & 1.8498 \\
			RDP   & 0.6876 & 0.7262 & 0.4527 & 0.5327 & 0.6618 & 0.7088 & 0.6944 & 0.7666 \\
			SED   & 2.0281 & 1.9634 & 1.8908 & 1.7857 & 19.7811 & 21.0457 & 21.0339 & 23.2532 \\
			TD-TR & \textbf{0.0604} & \textbf{0.0559} & \textbf{0.0441} & \textbf{0.0341} & \textbf{0.0584} & \textbf{0.0644} & \textbf{0.0629} & \textbf{0.0714} \\
			\cmidrule(lr){2-9}
			TDKC  & \multicolumn{8}{c}{0.0364}   \\
			ARDP  & \multicolumn{8}{c}{0.0034}  \\
			\bottomrule
		\end{tabular}
	}
\end{table}

Table \ref{tab_rt} shows that runtime scales consistently with algorithmic complexity across all three loitering categories. TD-TR, TDKC, and ARDP are the fastest by a wide margin. SED is by far the most expensive method, particularly in $\epsilon$-mode, where its cost rises to 8–23 s per trajectory because it must repeatedly re-solve a global positional-deviation optimization, and this cost grows quickly with trajectory density in Category 3. RDP  is nearly insensitive to $K$ or $\epsilon$, since threshold-based recursive splitting does not directly depend on the target point budget. CDS, by contrast, scales with $K$ as predicted by its $O(K\cdot N^2)$ complexity, and is slightly faster than RDP for category 1 and 2. For category 3, as $K$ grows lager, its runtime also increase quicker than RDP. PDS exhibits a flat runtime profile across $\epsilon$ and its runtime is comparable to RDP for category 1 and increase more rapidly for category 2 and 3. Overall, although dynamic programming is traditionally regarded as computationally intensive, our explicit formulation, precomputation, early stopping mechanism, and vectorized implementation ensure that the proposed framework remains viable for offline trajectory data processing.



}

\section{Conclusions and Discussions}
The length of the trajectory is crucial for downstream maritime applications, such as the estimation of fuel consumption and the auditing of carbon emissions. This study proposes a novel trajectory simplification method that explicitly optimizes trajectory length. We formulate the problem as two dual optimization tasks: Compression-Driven Simplification (CDS) and Precision-Driven Simplification (PDS), and achieve a unified provable optimal solution based on a dynamic programming framework. This framework allows for direct and flexible control over data volume and length approximation error. The proposed framework is validated through extensive experiments on real-world AIS data and compared with the state-of-the-art baseline methods. Extensive results demonstrate that our proposed method mathematically guarantees a {globally optimal allocation of retained points with respect to the trajectory-length preservation objective to maximally preserve trajectory length and geometric shape.} 

{

Although the proposed method demonstrates strong empirical effectiveness, there are some potential limitations. When processing long trajectories with high point densities, both the distance matrix and the dynamic programming table demand substantial memory resources. One trade-off way is to partition the trajectory into a few short ones and simplify each one independently. However, partitioning may sacrifice the ability to find a globally optimal solution. Furthermore, despite useful implementation-level acceleration techniques, the algorithm may still experience computational bottlenecks when scaled to massive, continuous AIS data streams over extended durations. In addition, the current objective function primarily prioritizes the preservation of cumulative trajectory length. Consequently, when encountering subtle, short-term navigational maneuvers, the algorithm may treat such local variations as redundant deviations, even though these localized patterns can be critical for maritime safety assessment.

}

Future work will explore strategies to accelerate the algorithm while maintaining approximation accuracy. Furthermore, although the current method focuses solely on the preservation of voyage length, the framework can be extended to multi-objective optimization problems by incorporating additional information regarding ship dynamics, thereby promoting the capability of digitalization of ship industry.



 \bibliographystyle{elsarticle-num-names} 
 \bibliography{cas-refs}

\appendix

\section{EEOI Error Calculation}
The core calculation of EEOI relies on the ratio of total $CO_{2}$ emissions to the transport work performed, which is defined as the product of cargo capacity and the actual sailing distance. More specifically, the EEOI calculation is as follows:

\begin{equation}
	EEOI = \frac{\sum_{j} FC_{j} \times C_{F_{j}}}{m_{cargo} \times D}
\end{equation}
where $FC_{j}$ denotes the mass of the consumed fuel type $j$, $C_{F_{j}}$ represents the carbon-to-$CO_{2}$ conversion factor for fuel type $j$, $m_{cargo}$ indicates the cargo payload, and $D$ corresponds to the actual sailing distance (i.e., the trajectory length). In the experimental calculations, heavy fuel oil (HFO) was selected as the specific fuel type, given its status as the primary fuel in international ocean shipping. Correspondingly, the  error of EEOI introduced by simplified trajectory is defined as the relative percentage difference between the simplified and original ones:

\begin{equation}
	Error_{EEOI} = \frac{EEOI_{sim} - EEOI_{orig}}{EEOI_{orig}} \times 100\%
\end{equation}
where $EEOI_{sim}$ and $EEOI_{orig}$ represent the EEOI values calculated using the simplified trajectory length and the original trajectory length, respectively.

\section{Full Simplified Trajectories for Case Studies}
\begin{figure}[h]
	\centering
	\begin{subfigure}{0.48\textwidth}
		\includegraphics[width=\linewidth]{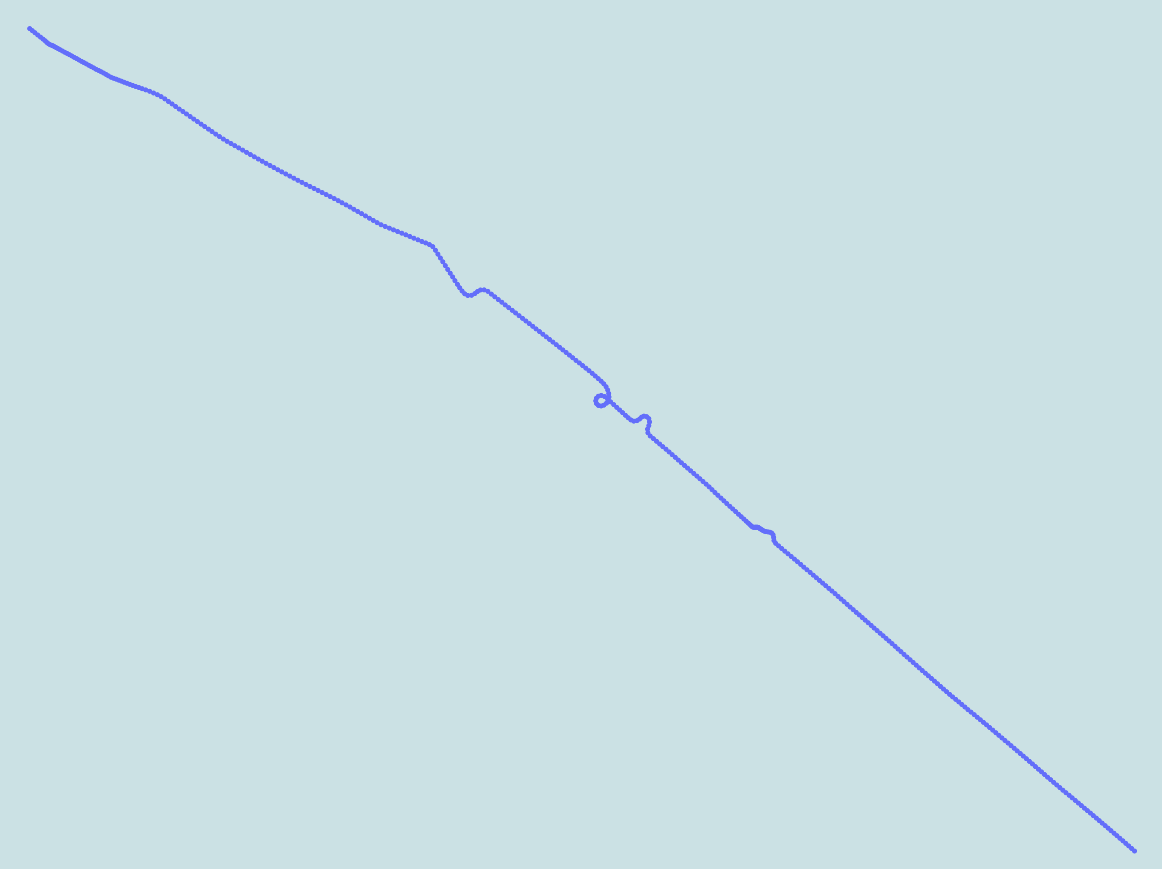}
		\caption{Original trajectory}
	\end{subfigure}
	\hfill
	\begin{subfigure}{0.48\textwidth}
		\includegraphics[width=\linewidth]{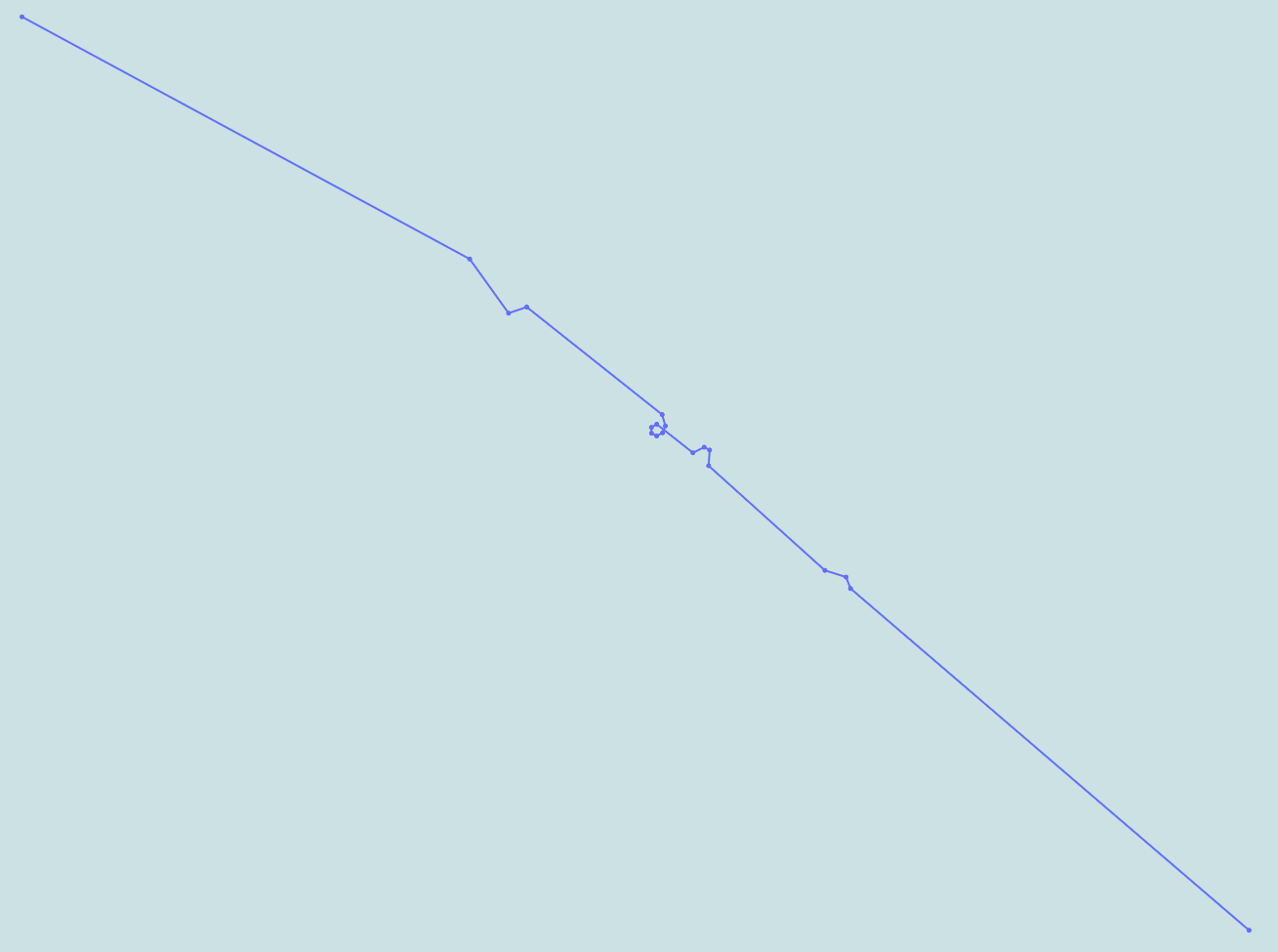}
		\caption{CDS ($\epsilon=0.6321$)}
	\end{subfigure}
	\hfill
	\begin{subfigure}{0.48\textwidth}
		\includegraphics[width=\linewidth]{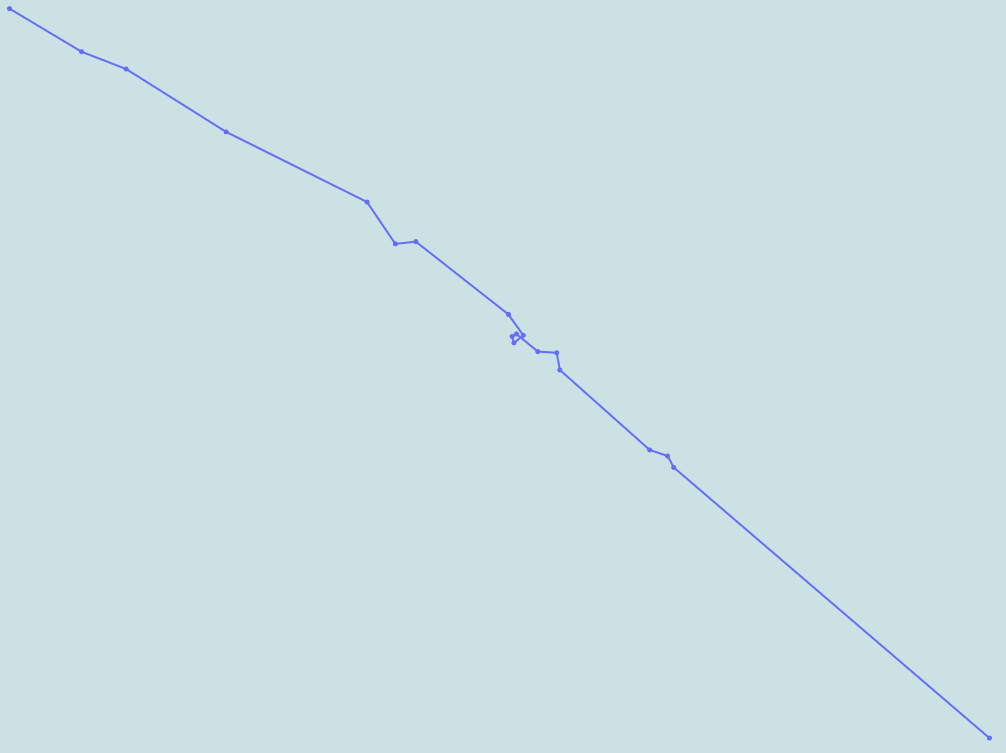}
		\caption{RDP ($\epsilon=1.3995$)}
	\end{subfigure}
	\hfill
	\begin{subfigure}{0.48\textwidth}
		\includegraphics[width=\linewidth]{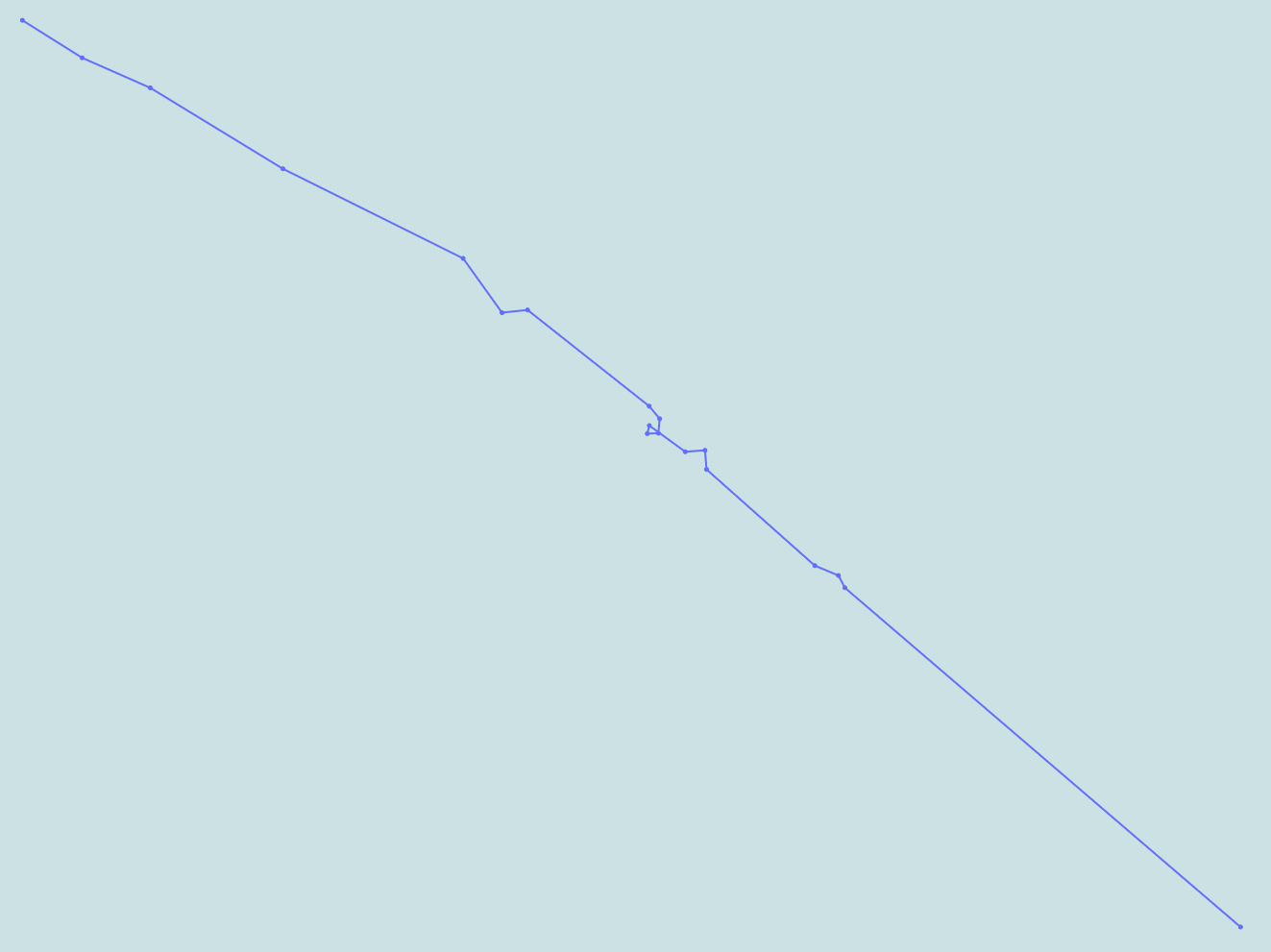}
		\caption{TD-TR ($\epsilon=1.3730$)}
	\end{subfigure}
	\hfill
	\begin{subfigure}{0.48\textwidth}
		\includegraphics[width=\linewidth]{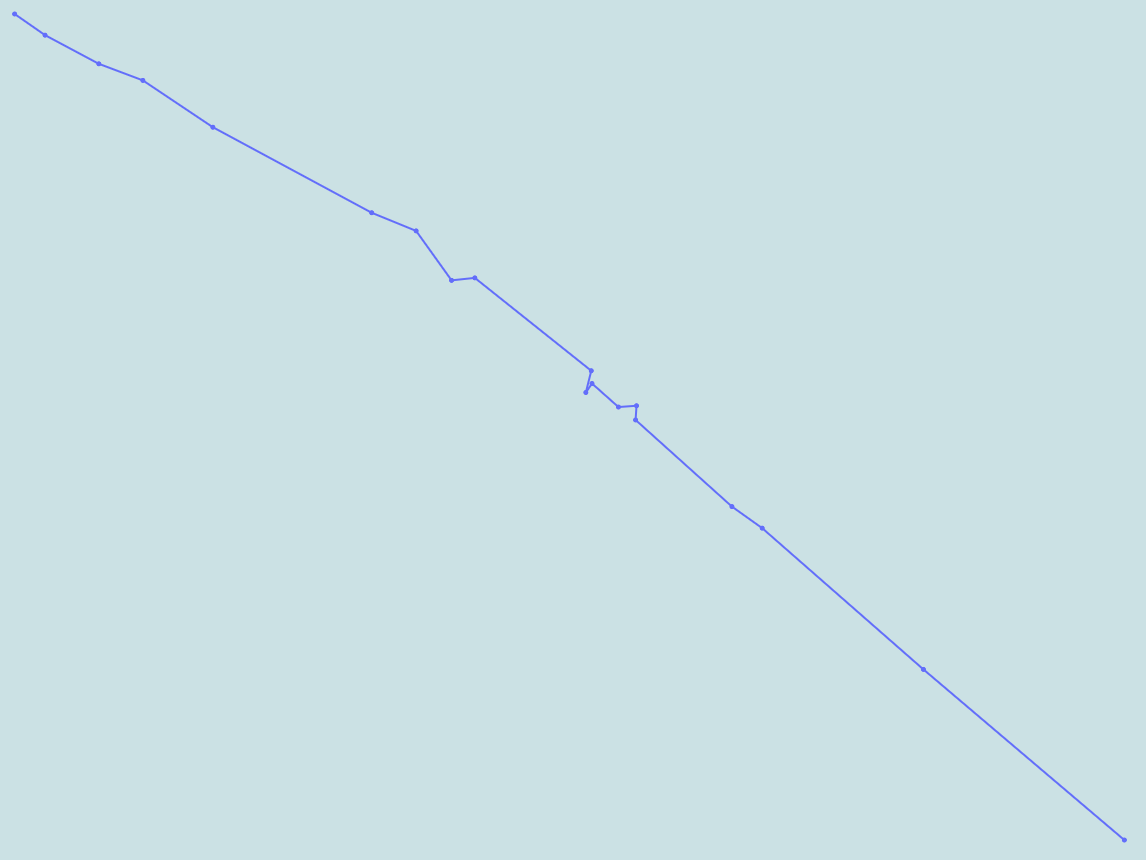}
		\caption{SED ($\epsilon=2.4278$)}
	\end{subfigure}
	
	\caption{Simplification results of case 1}
	\label{app_fig1}
\end{figure}

\begin{figure}[h]
	\centering
	\begin{subfigure}{0.48\textwidth}
		\includegraphics[width=\linewidth]{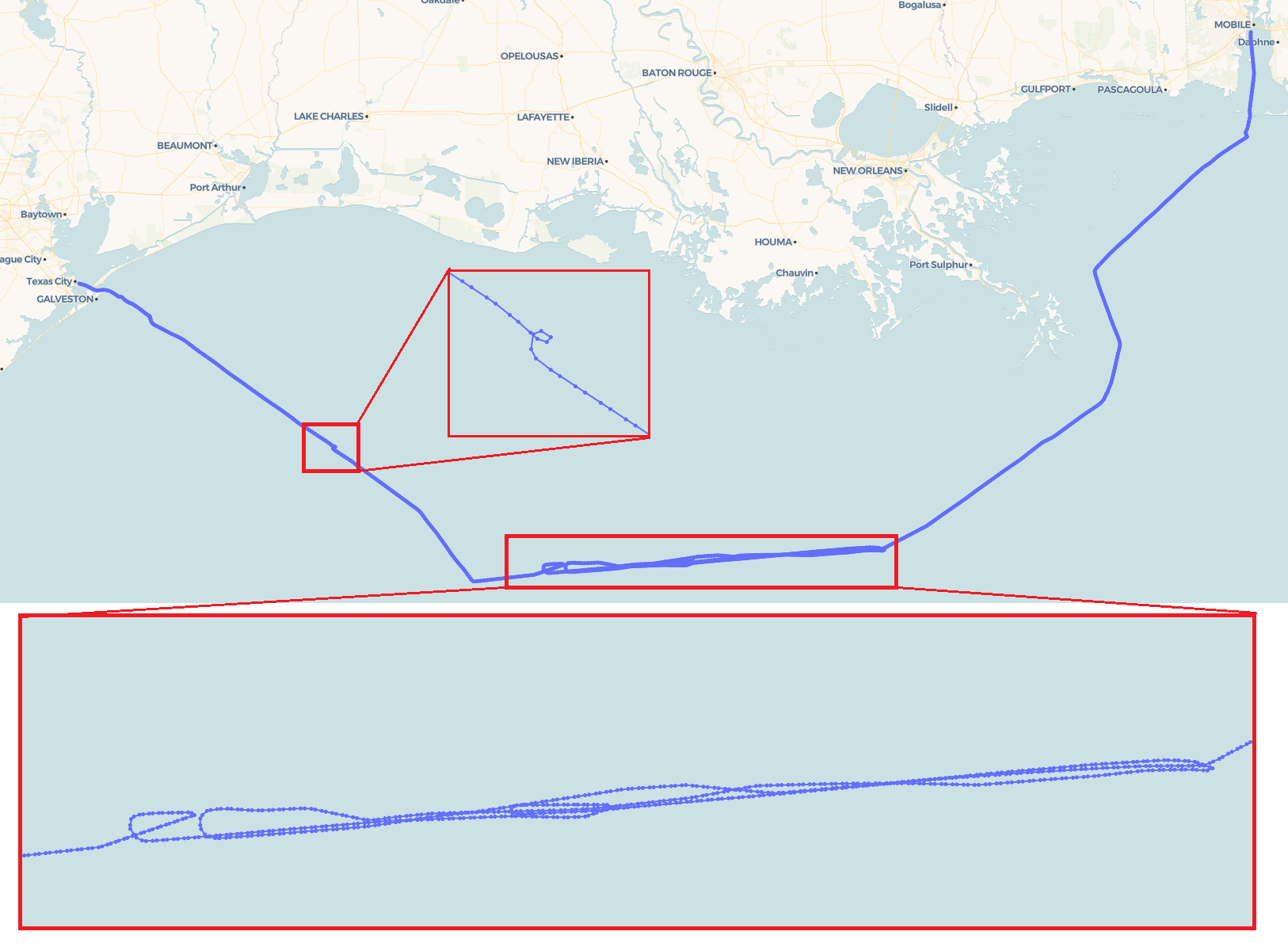}
		\caption{Original trajectory}
	\end{subfigure}
	\hfill
	\begin{subfigure}{0.48\textwidth}
		\includegraphics[width=\linewidth]{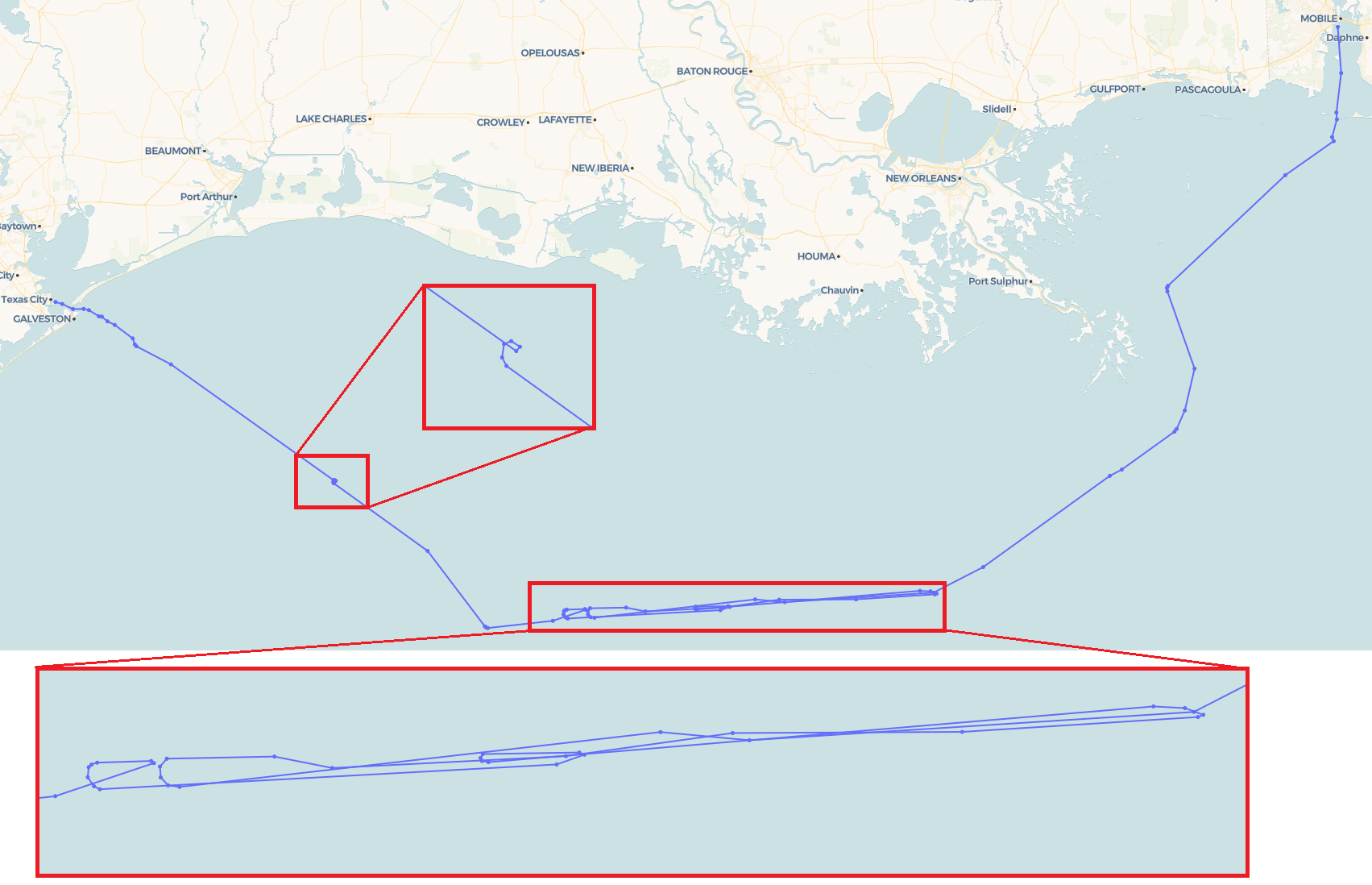}
		\caption{CDS ($\epsilon=0.5586$)}
	\end{subfigure}
		\hfill
	\begin{subfigure}{0.48\textwidth}
		\includegraphics[width=\linewidth]{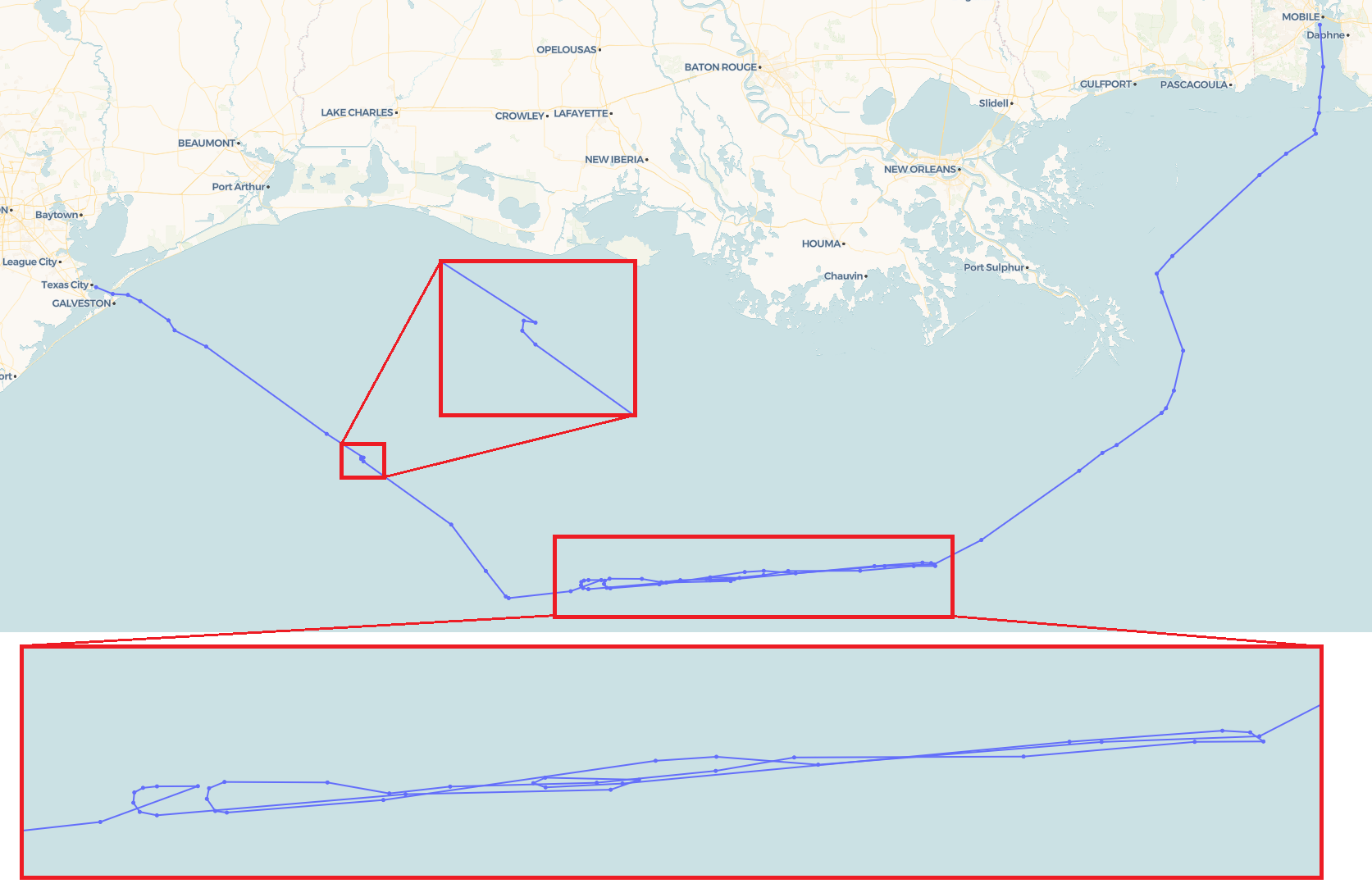}
		\caption{RDP ($\epsilon=2.8790$)}
	\end{subfigure}
		\hfill
	\begin{subfigure}{0.48\textwidth}
		\includegraphics[width=\linewidth]{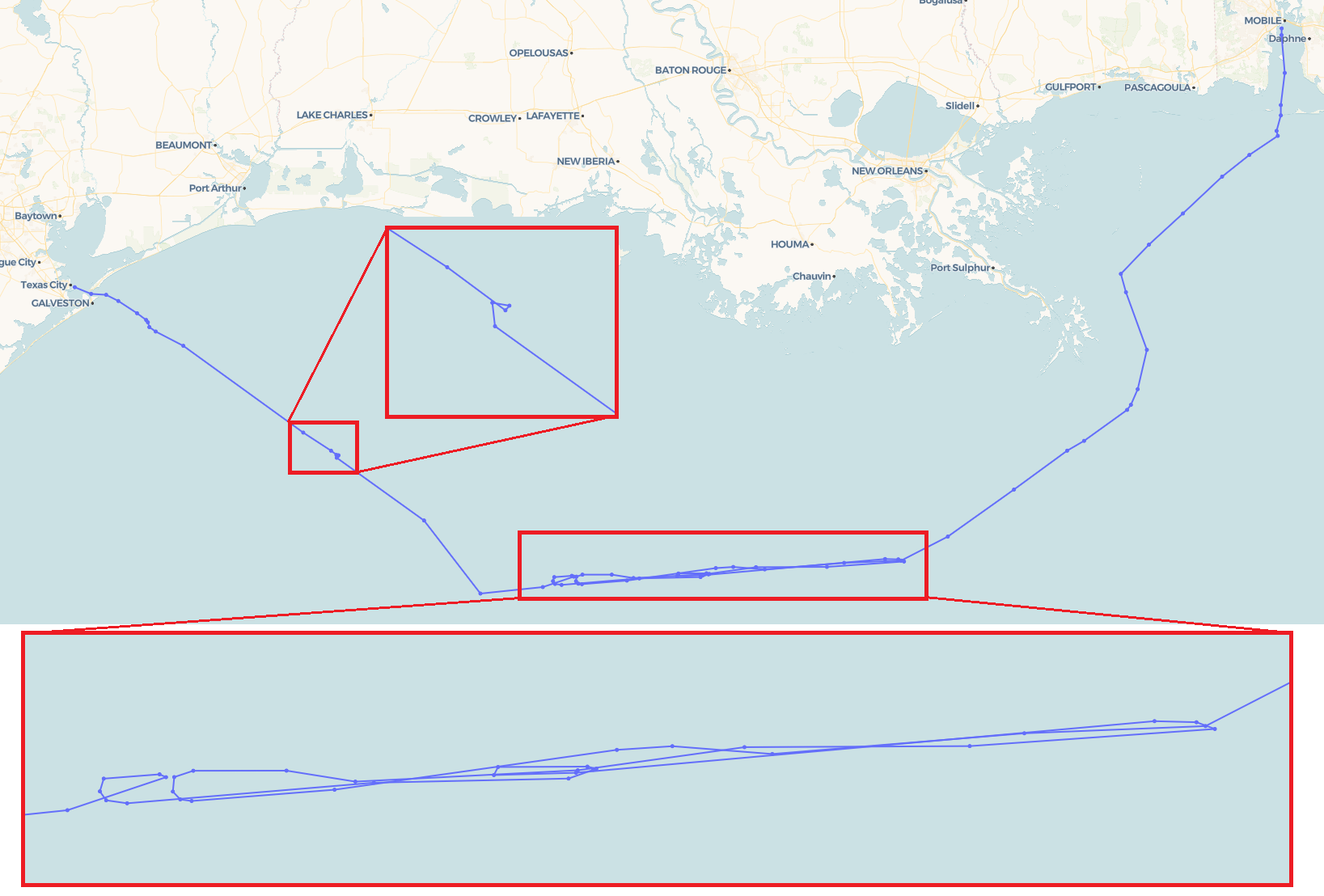}
		\caption{TD-TR ($\epsilon=1.4431$)}
	\end{subfigure}
		\hfill
	\begin{subfigure}{0.48\textwidth}
		\includegraphics[width=\linewidth]{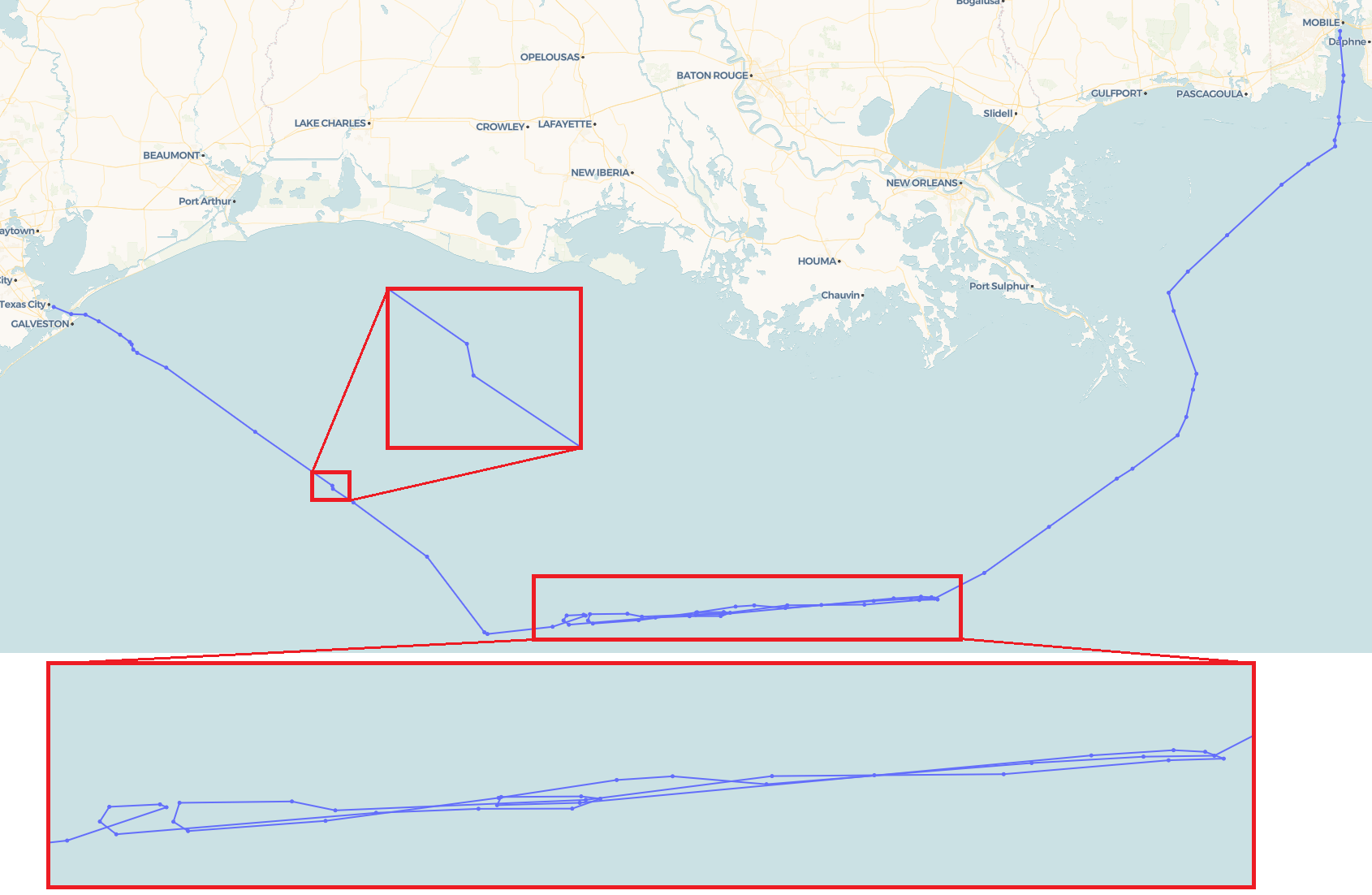}
		\caption{SED ($\epsilon=3.2783$)}
	\end{subfigure}
	
	\caption{Simplification results of case 2}
	\label{app_fig2}
\end{figure}

The full simplified trajectories of case 1 and case 2 are displayed in figure \ref{app_fig1} and \ref{app_fig2}, respectively.  We can see that CDS achieves significant lower length error than the baseline algorithms, meanwhile it  keeps better  geometric shape (particularly the small loops).

%
%
%





\end{document}